\documentclass[aps,prb,reprint,superscriptaddress,letterpaper,showkeys,floatfix]{revtex4-2}

\usepackage{color}
\usepackage{soul}
\usepackage{setspace}
\usepackage{gensymb}
\usepackage{graphicx}
\usepackage{epstopdf}
\usepackage{amssymb}
\usepackage[colorlinks=true,linkcolor=blue]{hyperref}
\usepackage{array}
\usepackage{bm}
\usepackage{float}
\usepackage{amsmath}
\usepackage{siunitx}
\DeclareSIUnit{\at}{at.\,\%}
\usepackage{multirow}
\usepackage[capitalize]{cleveref}
\usepackage{todonotes}
\usepackage{booktabs}
\usepackage{adjustbox}
\usepackage{ulem}
\usepackage{todonotes}

\newcommand{\phiMS}{\ensuremath{\Phi_\text{MS}}}
\newcommand{\Happl}{\ensuremath{H_\text{appl}}}
\newcommand{\HIP}{\ensuremath{H_\text{IP}}}
\newcommand{\HOOP}{\ensuremath{H_\text{OOP}}}
\newcommand{\bFe}{\ensuremath{  \text{bcc Fe} }}
\newcommand{\hFe}{\ensuremath{  \text{hcp Fe} }}
\newcommand{\hFeRu}{\ensuremath{ \text{hcp FeRu} }}
\newcommand{\bFeRu}{\ensuremath{ \text{bcc FeRu} }}
\newcommand{\xFe}{\ensuremath{ \mathrm{Fe}_{100-x}\mathrm{Ru}_{x} }}
\newcommand{\xFed}[1][]{\ensuremath{ \mathrm{Ta}|\mathrm{Ru}|\mathrm{Fe}_{100-x}\mathrm{Ru}_{x}(#1)|\mathrm{Ru}   }}
\newcommand{\Fed}[3]{\ensuremath{ \mathrm{Ta}|\mathrm{Ru}|\mathrm{Fe}_{#1}\mathrm{Ru}_{#2}(#3)|\mathrm{Ru}   }}

\begin{document}
	
        \title{Transition from ferromagnetic to noncollinear to paramagnetic state with increasing Ru concentration in FeRu films}

	\author{Juliana Lisik}
	\email{jbesler@sfu.ca}
    \author{Manuel Rojas}
    \email{mrojas@sfu.ca}
    \author{Spencer Myrtle}
	\affiliation{Simon Fraser University, 8888 University Drive, Burnaby, British Columbia V5A 1S6, Canada\looseness=-1}
	\author{Dominic H. Ryan}
	\email{dhryan@physics.mcgill.ca}
	\affiliation{Physics Department and Centre for the Physics of Materials, McGill University, 3600 University Street, Montreal, Quebec, H3A 2T8, Canada\looseness=-1}
	\author{Ren\'{e}~H\"{u}bner} 
	\affiliation{Institute of Ion Beam Physics and Materials Research, Helmholtz-Zentrum Dresden-Rossendorf, Bautzner Landstraße 400, 01328 Dresden, Germany\looseness=-1}
	\author{Pavlo Omelchenko}
        \affiliation{Simon Fraser University, 8888 University Drive, Burnaby, British Columbia V5A 1S6, Canada\looseness=-1}
	\author{Claas Abert}
        \author{Amil Ducevic}
        \author{Dieter Suess}
        \affiliation{Physics of Functional Materials, Faculty of Physics, University of Vienna, Kolingasse 14–16, 1090 Vienna, Austria\looseness=-1}
	\author{Ivan Soldatov}
        \author{Rudolf Schaefer}
        \affiliation{Leibniz-Institut für Festkörper- und Werkstoffforschung Dresden, Germany\looseness=-1}
        \author{Johannes Seyd}
        \author{Manfred Albrecht}
        \affiliation{Institute of Physics, University of Augsburg, 86135 Augsburg, Germany}
	\author{Erol Girt}
	\email{egirt@sfu.ca}
	\affiliation{Simon Fraser University, 8888 University Drive, Burnaby, British Columbia V5A 1S6, Canada\looseness=-1}

	\date{\today}

	\begin{abstract}

The structural and magnetic properties of sputter-deposited Fe$_{100-x}$Ru$_x$ films were studied for $x \leq 50$. The crystal structure of Fe$_{100-x}$Ru$_x$ is shown to be predominantly body-centered cubic for $x<13$ and to undergo a gradual transition to hexagonal close-packed in the concentration range $13 \lesssim x \lesssim 20$. Magnetic measurements indicate that the addition of Ru induces a noncollinear magnetic order in the body-centered cubic FeRu alloys, while the hexagonal close-packed FeRu alloys exhibit paramagnetic behavior. Increasing the Ru concentration in body-centered cubic FeRu alloys decreases the size of magnetic domains, approaching the size of magnetic grains. A simple atomistic model was used to show that antiferromagnetic coupling of Fe atoms across Ru atoms can be responsible for inducing noncollinear order in the FeRu cubic structures. Magnetic multilayer structures used in thin-film magnetic devices make extensive use of both Fe and Ru layers. Our results reveal that the presence of even a small amount of Ru in Fe influences the magnetic order of Fe, which could impact the performance of these devices.
	
 	\end{abstract}

	\maketitle

	\section{Introduction}

Iron is the most abundant element in the Earth by mass, composing about 80\% of Earth’s core; furthermore, it is the fourth most abundant element in Earth's crust \cite{frey2012the}. The structural, electrical, and magnetic properties of iron and its alloys are at the root of many applications. Of the three elements that have ferromagnetic alignment above room temperature (iron, cobalt, and nickel), iron has the largest saturation magnetization, making iron alloys a crucial component of magnetic devices (e.g. hard and soft magnets, magnetic sensors, and recording media). However, despite centuries of research, understanding the phase diagram and magnetic properties of iron alloys remains an active research topic.

The unary phase diagram of Fe consists of three crystal structures that meet at a triple point located at approximately \SI{477}{\degree C} and \SI{11}{GPa}.
At standard temperature and pressure, Fe is body-centered cubic (bcc), which is known as $\alpha$-Fe, and has well-known ferromagnetic order.
The crystal structure transitions from bcc to face-centered cubic (fcc) at high temperatures.
At pressures above \SI{11}{GPa}, the crystal structure of Fe transitions from bcc to hexagonal close-packed (hcp), which is known as $\epsilon$-Fe. The \hFe{} phase is stable for pressures and temperatures corresponding to the conditions of Earth's core \cite{takahashi1964high}.
At the transition from bcc to \hFe, there is a rapid decrease in magnetic moment \cite{Mathon2004FehcpTobcc}.
However, unlike \bFe, the magnetic state of \hFe{} is not well understood.
Some experimental studies find that \hFe{}  does not appear to have any long-range magnetic order, as determined from low-temperature M\"ossbauer measurements \cite{PapandrewFehcpMag,Nasu2002}, X-ray magnetic circular dichroism (XMCD) \cite{Mathon2004FehcpTobcc}, and neutron powder diffraction \cite{Lebert2019FeModel}. X-ray emission spectroscopy, on the other hand, detects a localized magnetic moment for \hFe{} between about 12 and \SI{40}{GPa} \cite{Monza2011,Lebert2019FeModel} with a magnitude of $0.74 \mu_\text{B}$ at \SI{20}{GPa} \cite{Lebert2019FeModel}. More recently, a ground state of alternating antiferromagnetic and nonmagnetic atomic layers was used to explain the conflicting experimental observations discussed above \cite{Lebert2019FeModel}.

Thin-film magnetic devices make extensive use of not only Fe but also Ru layers \cite{gallagher2006development,chen2017high,guedes2006study}. Thus, it is important to study how the presence of Ru in Fe, which can occur during deposition or annealing, affects the magnetic properties of Fe.
When Fe is alloyed with Ru, not only the bcc phase but also the hcp phase can be stabilized at standard temperature and pressure. The bcc phase of FeRu has been seen for up to \SI{4}{\at} of Ru in Fe \cite{Swartzendruber1983}. Recent studies suggest that FeRu is ferromagnetic in this phase \cite{Pottker2004}, as is pure Fe. Additionally, M\"ossbauer studies of bulk bcc FeRu argue for the presence of spin and charge density waves in the vicinity of the Ru impurity in Fe \cite{Blachowski2006SpinDensity}. As more Ru is added, the hcp phase appears: it has been seen above \SI{23}{\at} of Ru in the equilibrium phase diagram \cite{Swartzendruber1983}.
There are many parallels between high-pressure \hFe{} and hcp FeRu; however, theoretical and experimental studies of FeRu are scarce. Similar to high-pressure Fe studies, during the transition from bcc to hcp, the magnetization of FeRu collapses. The magnetic structure of the hcp FeRu alloy remains a debated topic. Early M\"ossbauer studies did not observe any hyperfine structure, but the existence of an antiferromagnetic phase with a magnetic moment of about $0.1 \mu_\text{B}$ was estimated \cite{ohno1971antiferromagnetism,pearson1979mossbauer}. More recent M\"ossbauer studies suggest that hcp FeRu is paramagnetic \cite{Pottker2004}. Finally, neutron diffraction studies of bulk hcp Fe$_{71}$Ru$_{29}$ reported an incommensurate antiferromagnetic state with a N\'eel temperature of \SI{124}{K} and a low-temperature Fe magnetic moment of about $\mu_\text{B}$ \cite{PetrilloND2018}. Theoretical work on the ground state of Fe$_{50}$Ru$_{50}$ identifies a type-II antiferromagnetic phase as the lowest energy state \cite{Marcus1996AFM2}. 

More recently, thin-film FeRu has been shown to mediate a large noncollinear magnetic coupling between two ferromagnetic layers \cite{nunn2020control}. The coupling angle and strength was controlled by the concentration of Ru within the FeRu alloy. Controlling the coupling angle is important for the design of magnetic thin-film devices, as the optimal structure almost always requires noncollinear alignment between at least two adjacent ferromagnetic films \cite{fullerton2016spintronics,sbiaa2013magnetization,matsumoto2015spin,zhou2008spin}. At present, it is not clear what role the bcc and hcp structures within the FeRu alloy play in the observed noncollinear coupling.

In this paper, we investigate the structural and magnetic properties of FeRu films over a composition range from 0 to \SI{50}{\at} of Ru in Fe. The emphasis is on the range of Ru concentrations from 0 and \SI{20}{\at}, within which the transition from bcc to hcp crystal structure occurs.
Structural properties are explored with X-ray diffraction and transmission electron microscopy, and magnetic properties are explored with M\"ossbauer spectroscopy, vibrating sample magnetometry, magneto-optical Kerr effect microscopy, and Lorentz transmission electron microscopy.
Our results show that, as Ru is added to ferromagnetic Fe films, the magnetic behaviour of the films transitions to noncollinear and then to paramagnetic.
Furthermore, we argue that noncollinear coupling is a result of the competition between ferromagnetic Fe-Fe coupling and antiferromagnetic coupling between Fe atoms separated by Ru atoms.

	\section{Experimental Details}
	
The studied films, Ta(3.5)$|$Ru(3.5)$|$Fe$_{100-x}$Ru$_{x}$($d$)$|$ Ru(3.5),
were deposited with radio-frequency magnetron sputtering on (100) Si, Si$_{3}$N$_{4}$ membranes, and Kapton substrates at room temperature and an argon pressure below \SI{2}{mTorr}. 
In these films, the numbers in parentheses indicate the layer thicknesses in nm, $x$ is the atomic concentration of Ru in Fe$_{100-x}$Ru$_{x}$, and $d=20$ or $\SI{100}{nm}$ is the thickness of the FeRu layer. Si substrates were used for all studies except for transmission M\"ossbauer spectroscopy, which requires Kapton substrates to maximize the signal.
Throughout this paper, we will refer to Ta(3.5)$|$Ru(3.5)$|$Ru$_{100-x}$Fe$_{x}$($d$)$|$Ru(3.5) films as \xFed[d], as the thicknesses of the seed and capping layers were kept the same for all samples. 
The Ta seed layer is deposited to induce the $\langle$0001$\rangle$ growth orientation of the bottom Ru layer, and the top Ru film is used to protect the Ru$_{100-x}$Fe$_{x}$ layer from oxidation.
Selected films were annealed at \SI{400}{\celsius} for 1~hour to study the thermal stability of the phase structure.

Prior to deposition, (100) Si substrates were cleaned with the standard RCA SC-1 process to remove particles and organic contaminants. Kapton substrates were cleaned with isopropyl alcohol and ethanol. The substrates were subsequently rinsed in deionized water and dried. Clean substrates were first placed in a load lock chamber, which is evacuated to about 5$\times 10^{-7}$ Torr, and then transferred, without breaking the vacuum, to a process chamber with base pressure below 5$\times 10^{-8}$ Torr for deposition. The films were deposited from three elemental \SI{5.08}{cm} (\SI{2}{inch}) diameter targets of Ta, Ru, and Fe.
The target-to-substrate distance is approximately \SI{20}{cm}. The substrate holder rotates during deposition to ensure thickness and composition uniformity of the deposited films across the substrate surface. The entire sputter process is computer-controlled.
For the Ta and pure Ru layers, the sputtering rates were 0.26 and \SI{0.22}{\angstrom/s}, respectively. To deposit the Fe$_{100-x}$Ru$_{x}$ layer, the sputtering rate for Fe was varied from \SI{0.30}{\angstrom/s} ($x=50$) to \SI{0.59}{\angstrom/s} ($x=4$), and the sputtering rate for Ru was varied from \SI{0.35}{\angstrom/s} ($x=50$) to \SI{0.03}{\angstrom/s} ($x=4$). The pure Fe layer was sputtered with a rate of \SI{0.36}{\angstrom/s}.
The sputtering rates were inferred from X-ray diffraction (XRD) measurements performed using CuK$_\alpha$ radiation. Low-angle out-of-plane $\theta$-$2\theta$ (reflectivity) measurements of single layers of each material or of Ru(2)$|$X$|$Ru(2) multilayers (X = Ta, Fe, and FeRu) were fit with X$'$PERT reflectivity software from Panalytical.

Structural characterization of the deposited layers was carried out by XRD using out-of-plane $\theta$-$2\theta$, in-plane $\theta$-$2\theta$, and rocking curve measurements.
For the out-of-plane and in-plane $\theta$-$2\theta$ measurements, the angle between the incident X-ray beam and the film surface, $\theta_1$, is the same as the angle between the reflected X-ray beam and the film surface, $\theta_2$, i.e., $\theta_1=\theta_2=\theta$. For the rocking curve measurement, $\theta_1+\theta_2$ is kept constant and the sample is rocked by angle $\omega$ so that $\theta_1$ varies from $\theta-\omega/2$ to $\theta+\omega/2$.
The scattering wave vector is perpendicular to the film surface for the out-of-plane $\theta$-$2\theta$ measurement and varies around the perpendicular direction for the rocking curve measurement.
For the in-plane $\theta$-$2\theta$ measurement, however, the scattering wave vector is almost parallel to the film surface, at an angle of $0.5\degree$.

To further characterize the film microstructure, bright-field and high-resolution transmission electron microscopy (HRTEM) imaging were performed on \xFed[20] ($x=13$ and 17) and Ta$|$Ru$|$Fe(100)$|$Ru films on an image-C$_\text{s}$-corrected Titan 80-300 microscope (FEI) operated at an accelerating voltage of \SI{300}{kV}.
Scanning transmission electron microscopy (STEM) coupled with spectrum imaging analysis based on 
energy-dispersive X-ray spectroscopy (EDXS) was performed at \SI{200}{kV} with a Talos F200X microscope equipped with a Super-X EDX detector system (FEI).
The HRTEM images were further examined by fast Fourier transform (FFT) analysis to detect the spatial distribution of different lattice structures.
Prior to TEM analysis, the specimen, mounted in a high-visibility low-background holder, was placed into a Model 1020 Plasma Cleaner (Fischione) for \SI{10}{s} to remove possible organic contamination.

The $^{57}$Fe M\"ossbauer spectroscopy measurements were carried out on \xFed[100] ($0\leq x\leq 50$), Ta$|$Fe(100)$|$Ta, and Fe(100) films.
For each film, one deposition was performed, sputtering on Kapton film taped to an \SI{8}{in}-diameter Si substrate for M\"ossbauer measurements and simultaneously sputtering on \SI{1}{in}-diameter Si substrates for magnetometry measurements, which are discussed in the next paragraph. For conversion electron M\"ossbauer spectroscopy (CEMS) measurements, $\sim$\SI{50}{mm}-diameter Kapton film samples were used (removed from the Si substrate), and for transmission M\"ossbauer spectroscopy (TMS) measurements, stacks of fifteen \SI{15}{mm} square Kapton film samples (also removed from the Si substrate) were made. This is because TMS measurements require thicker samples in order to obtain a useful signal-to-noise ratio.
M\"ossbauer measurements were performed using a \SI{1.8}{GBq} $^{57}$Co\underline{Rh} source driven in constant acceleration mode and calibrated using a thin \bFe{} foil. Isomer shifts are quoted relative to the centroid of the \bFe{} spectrum.
Room-temperature CEMS measurements were collected using a conventional homemade detector. The $\sim$\SI{50}{mm}-diameter Kapton film sample is mounted inside the detector with a flowing 90\%\,He + 10\%\,CH$_4$ gas mixture used to detect the backscattered electrons.
All films were measured at room temperature, but the \xFed[100] ($x=18.5$) film was also measured with TMS at \SI{9.7}{K}. Low temperatures were obtained by mounting the stacked samples in a vibration-isolated closed-cycle helium refrigerator.
All films were measured at zero external magnetic field, but Fe(100) and \xFed[100] ($x=4$ and $8$) were also measured in a \SI{0.166}{T} magnetic field created by permanent magnets to remove magnetic domains. The magnetic components of the spectra were fitted using an assumed Gaussian distribution of hyperfine fields with independent widths to the high and low sides of the peak of the distribution. This form is needed to fit the clearly asymmetric magnetic patterns observed. The paramagnetic components were fitted variously as: symmetric doublets with a small, unresolved, quadrupole splitting; a Gaussian distribution of quadrupole splittings; and, where the two lines in the doublet are clearly of unequal intensity, two independent lines.

The field dependence of the magnetization, $M(H)$, of \xFed[d] ($d=20$ and \SI{100}{nm}), Ta$|$Fe(100)$|$Ta, and Fe(100) films was measured using a vibration sample magnetometer (VSM)  and superconducting quantum interference device (SQUID) VSM in magnetic fields of up to \SI{7}{T} and at temperatures of 295 and \SI{5}{K}. 
Measurements were performed with the magnetic field applied parallel as well as perpendicular to the film surface.

A wide-field magneto-optical Kerr microscope was used to image magnetic domain structure in the Ta$|$Ru$|$Fe$_{96}$Ru$_{4}(100)$$|$Ru film. An external magnetic field was applied parallel to the film plane and the magnetic domain structure was observed, utilizing the longitudinal magneto-optical Kerr effect (MOKE) with two complementary sensitivity directions: along and transverse to the applied field.

The magnetic domain morphology of Ta$|$Ru$|$Fe$_{92}$Ru$_{8}$(20)$|$Ru was imaged by Lorentz transmission electron microscopy (LTEM) at room temperature using a JEOL NEOARM-200F system operated at \SI{200}{keV} beam energy in the Fresnel mode with an underfocus of \SI{-1}{mm}. Magnetic fields up to \SI{2}{\tesla} can be applied along the electron beam direction, which is typically perpendicular to the film plane. In order to apply an in-plane magnetic field in addition to the out-of-plane magnetic field, the sample must be tilted by a tilt angle $\theta$ with respect to the film normal. The sample for LTEM was prepared by magnetron sputter deposition on a \SI{30}{nm}-thick Si$_{3}$N$_{4}$ membrane, and images were acquired with a Gatan OneView camera.

	\section{Film Microstructure}
	
	\subsection{X-ray diffraction}
	\label{sec:XRD}
 
\begin{figure}[tbph]
			\centering			\includegraphics[width=8.6cm]{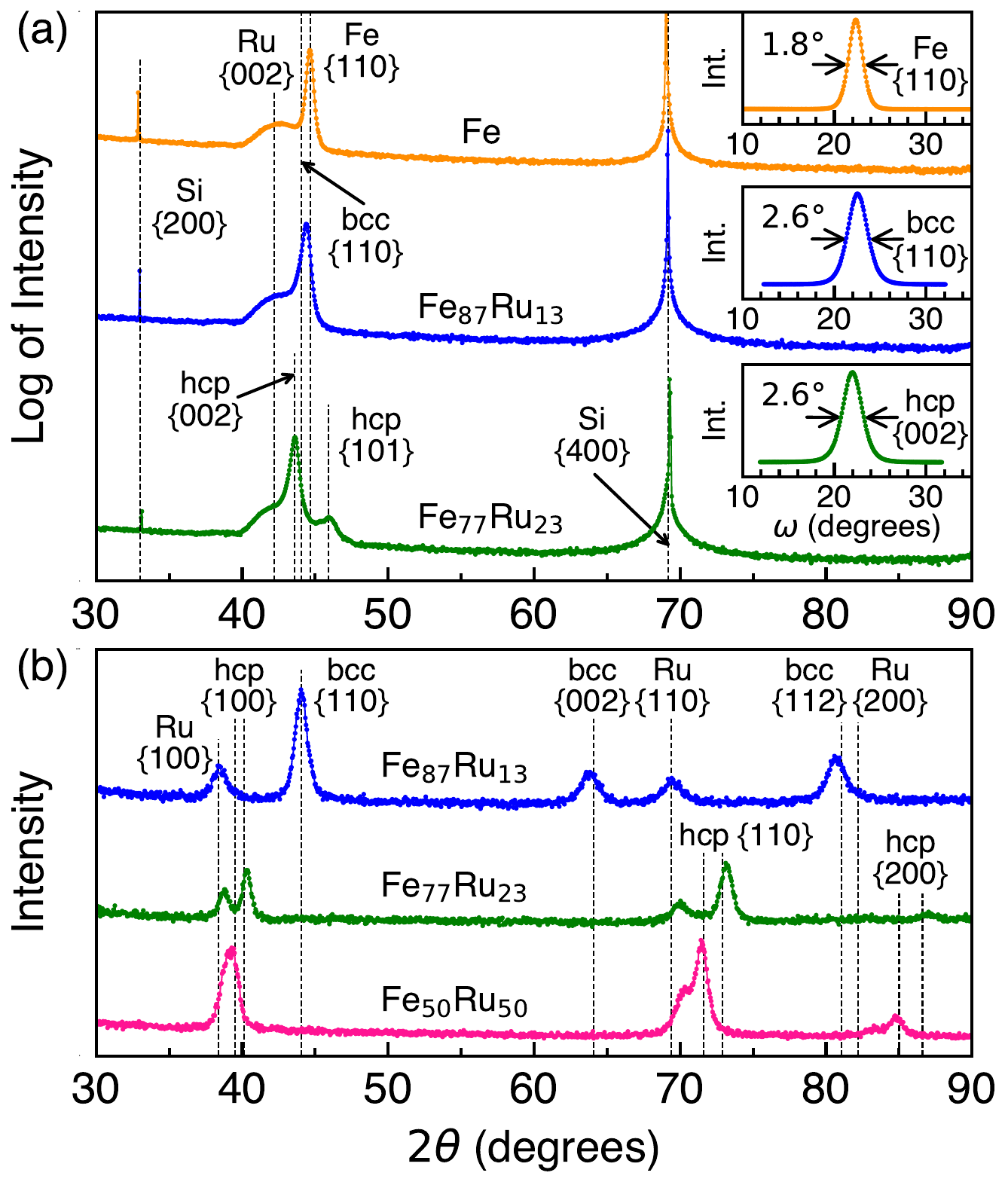}
            \caption{(a) Out-of-plane and (b) in-plane $\theta$-$2\theta$ measurements of as-deposited Ta(3.5)$|$Ru(3.5)$|$Fe$_{100-x}$Ru$_x$(100)$|$Ru(3.5) for $x=0$, 13, 23, and 50. The inset plots in (a) represent the rocking curve measurements of these samples and include the full-width-at-half-maximum.
            Dashed lines are placed at the expected positions of diffraction peaks.
            Those labelled Fe and Ru correspond to bcc Fe and hcp Ru, respectively, and those labelled hcp and bcc correspond to hcp Fe$_{100-x}$Ru$_x$ ($x=23$ or 50) and bcc Fe$_{87}$Ru$_{13}$, respectively.
            The Si \{200\} and \{400\} diffraction peaks in (a) are from the Si substrate.
            }
 		\label{fig:thickfilm}
 	\end{figure}

\begin{figure}[tbph]
			 \centering
    \includegraphics[width=8.6cm]{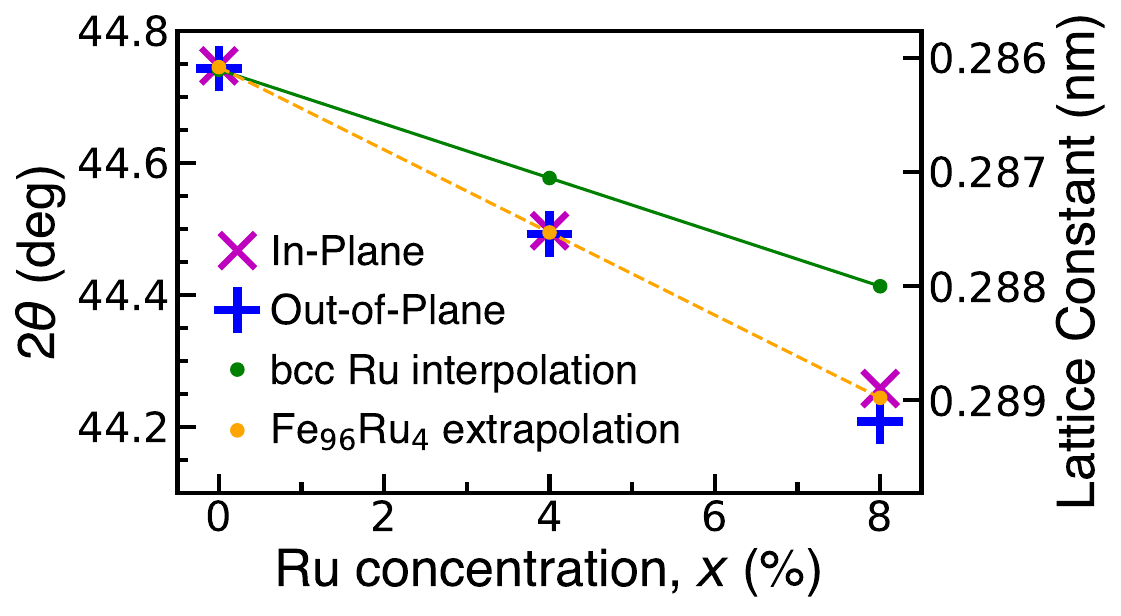}
            \caption{$2\theta$ value of the bcc \{110\} diffraction peak from in-plane and out-of-plane $\theta$-$2\theta$ XRD measurements of as-deposited Ta(3.5)$|$Ru(3.5)$|$Fe$_{100-x}$Ru$_x$(100)$|$Ru(3.5) ($x=0$, 4, and 8). The corresponding lattice constants of the bcc structure are shown on the secondary axis. The bcc Ru interpolation is a linear function using the measured lattice constants of our Fe film and the lattice constant of a hypothetical bcc unit cell of Ru atoms with radius \SI{.1359}{\nano\meter} \cite{huang2019}. The Fe$_{96}$Ru$_{4}$ extrapolation is a linear function using the measured lattice constants of our Fe and Fe$_{96}$Ru$_4$ films. The size of the markers corresponds to the size of the error bars.}
 		\label{fig:xraytrend}
 	\end{figure}

XRD measurements were performed on \xFed[$d$]{} films for $d=20$ and \SI{100}{nm}.
The out-of-plane and in-plane $\theta$-$2\theta$ measurements of as-deposited \xFed[100] for $x=0$, 13, 23, and 50 in \cref{fig:thickfilm} reveal that the Ru layers have hcp lattice structure, while the lattice structure of Fe$_{100-x}$Ru$_x$ depends on the Ru concentration, $x$. For $x=23$ and 50, Fe$_{100-x}$Ru$_{x}$ has hcp lattice structure, and for $x=13$, Fe$_{100-x}$Ru$_{x}$ has bcc lattice structure with no hcp lattice structure visible at the precision of our XRD measurements.
We obtained this information by comparing the positions of the measured diffraction peaks with the expected positions of hcp and bcc diffraction peaks (represented by the dashed lines in \cref{fig:thickfilm}) from previous XRD data as well as our own.
Previous XRD data on hcp Ru and Fe$_{50}$Ru$_{50}$ \cite{ISCD5050} were used to determine the positions of the hcp Ru lines, the hcp Fe$_{50}$Ru$_{50}$ lines (in \cref{fig:thickfilm}(b) only), and the hcp Fe$_{77}$Ru$_{23}$ lines (by extrapolation). Our out-of-plane and in-plane XRD data on bcc Fe and Fe$_{96}$Ru$_{4}$ (\cref{fig:xraytrend}), which will be discussed later, were used to determine the positions of the bcc Fe line (in \cref{fig:thickfilm}(a) only) and the bcc Fe$_{87}$Ru$_{13}$ lines (by extrapolation).

From the out-of-plane $\theta$-$2\theta$ measurements in \cref{fig:thickfilm}(a) for $x=0$, 13 and 23, it is apparent that hcp Ru and Fe$_{77}$Ru$_{23}$ have a \{002\} texture and that bcc Fe and Fe$_{87}$Ru$_{13}$ have a \{110\} texture. The diffraction pattern of hcp Fe$_{77}$Ru$_{23}$ also reveals a \{101\} diffraction peak with an intensity two orders of magnitude smaller than the \{002\} diffraction peak.
The quality of the texture of the three films plotted in \cref{fig:thickfilm}(a) is determined by rocking curve measurements, which are displayed in the inset plots above the spectrum of each film. The \{110\} reflection was measured for bcc Fe and Fe$_{87}$Ru$_{13}$, while the \{002\} reflection was measured for hcp Fe$_{77}$Ru$_{23}$. These plots indicate that the full-width-at-half-maximum (FWHM) of the rocking curve is $2.6\degree$ for the \{110\} reflection of Fe$_{87}$Ru$_{13}$ and for the \{002\} reflection of Fe$_{77}$Ru$_{23}$, suggesting highly textured FeRu films. For Fe, the FWHM of the rocking curve is $1.8\degree$ for the \{110\} reflection, indicating a stronger texture compared to FeRu films.
\Cref{fig:thickfilm}(a) also shows Si \{400\} and a small trace of Si \{200\} reflections from the substrate. Si \{200\} is forbidden and could be a result of multiple diffraction of X-rays in the crystal \cite{zaumseil2015high}.

Our films are polycrystalline with grains textured along the growth direction and randomly oriented in the plane of the film. For this reason, in-plane $\theta$-$2\theta$ measurements can be used to detect reflections from a range of planes perpendicular to the film surface, while out-of-plane $\theta$-$2\theta$ measurements can only detect the planes along which the film is textured. For the in-plane X-ray measurements, the diffraction pattern remains unaffected by the direction of the incident beam direction, owing to the random in-plane orientation of the grains. In contrast, if the sample has a single crystal structure, the incident beam must be precisely oriented with respect to the crystal lattice to obtain the diffraction pattern for in-plane measurements. Consequently, in-plane X-ray measurements allow one to distinguish between textured polycrystalline and single crystal samples. The in-plane $\theta$-$2\theta$ measurements for $x=13$, 23, and 50 are presented in \cref{fig:thickfilm}(b). These measurements show that, for hcp Ru, hcp Fe$_{50}$Ru$_{50}$, and hcp Fe$_{77}$Ru$_{23}$, we observe reflections from the \{100\}, \{110\}, and \{200\} planes. For each of these families of planes, the expected diffraction peak positions for Ru, Fe$_{50}$Ru$_{50}$, and Fe$_{77}$Ru$_{23}$ are shown. For bcc Fe$_{87}$Ru$_{13}$, we observe reflections from the \{110\}, \{002\}, and \{112\} planes.
Since reflections are detected from multiple planes\textemdash a consequence of the random in-plane orientation of the grains\textemdash the in-plane $\theta$-$2\theta$ measurements are better suited to detect the phase structure in our films than the out-of-plane measurements.
\Cref{fig:thickfilm}(b) shows that, as the concentration of Ru in Fe$_{100-x}$Ru$_{x}$ decreases from 50 to \SI{23}{\at}, the \{100\}, \{110\}, and \{200\} hcp diffraction peaks shift to higher angles, indicating a decrease in the lattice parameters of the FeRu layer. This is expected, because Fe and Ru form a solid solution in this composition range and Fe atoms are smaller than Ru atoms. Our results indicate that Fe$_{100-x}$Ru$_{x}$ transitions from hcp to bcc as the concentration of Ru is further decreased from 23 to \SI{13}{\at}. These observations are consistent with previous structural FeRu studies \cite{qadri2007structural,qadri2009thermal,pearson1979mossbauer}.

\begin{figure}[tbph]
			\centering
			\includegraphics[width=8.6cm]{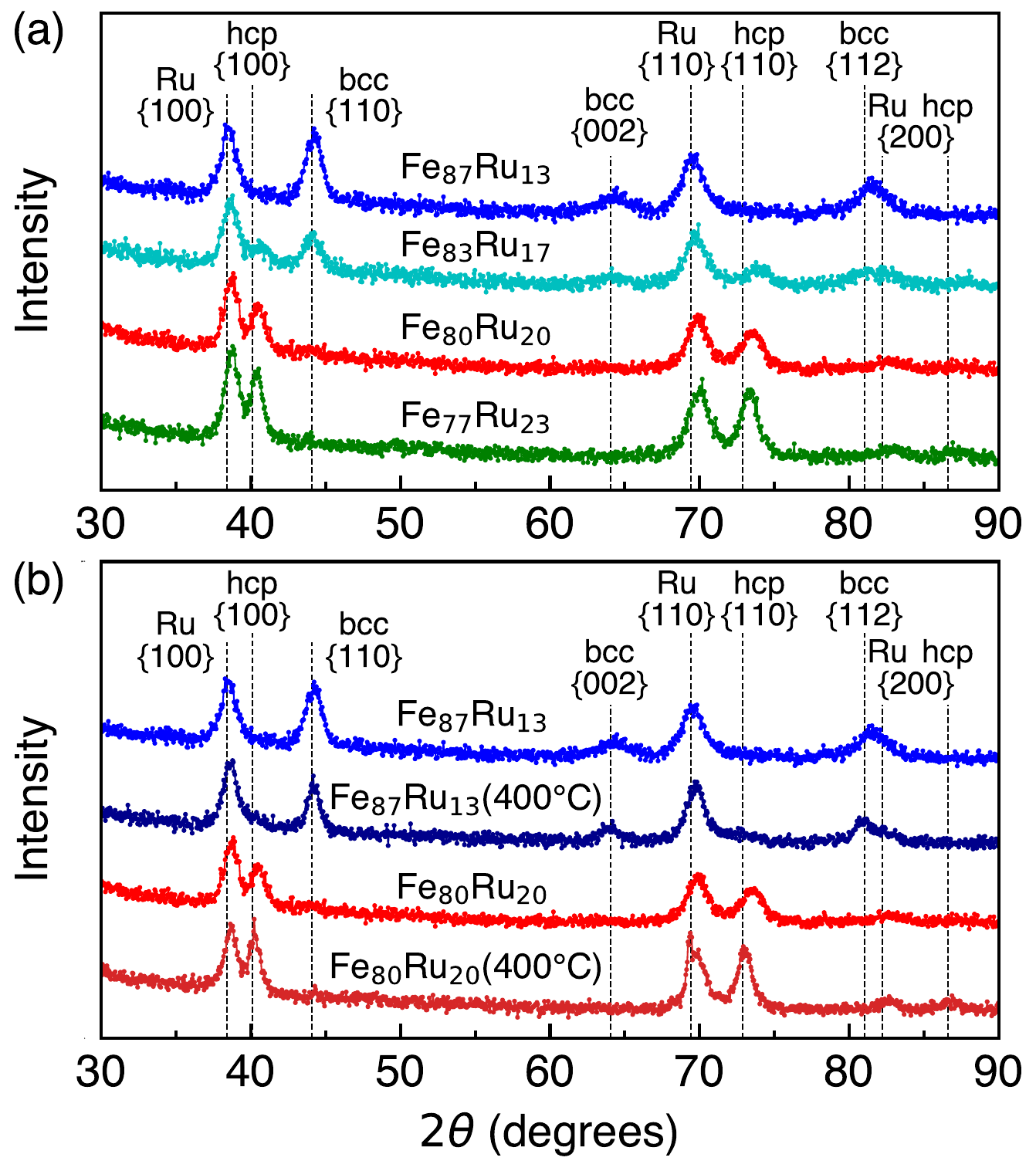}
            \caption{(a) In-plane $\theta$-$2\theta$ measurements of as-deposited Ta(3.5)$|$Ru(3.5)$|$Fe$_{100-x}$Ru$_x$(20)$|$Ru(3.5) for $x=13$, 17, 20, and 23. (b) In-plane $\theta$-$2\theta$ measurements before and after annealing at \SI{400}{\celsius} for $x=20$ and 13.
            Dashed lines are placed at the expected positions of diffraction peaks.
            Those labelled Ru correspond to hcp Ru, and those labelled hcp and bcc correspond to  hcp Fe$_{77}$Ru$_{23}$ and bcc Fe$_{87}$Ru$_{13}$, respectively.
            }
 		\label{fig:thinfilm}
 	\end{figure} 
  
In-plane $\theta$-$2\theta$ XRD measurements of \SI{20}{nm}-thick FeRu films, \xFed[20], are presented in \cref{fig:thinfilm}(a) for $x=13$, 17, 20, and 23. The expected diffraction peak positions are calculated in the same way as those in \cref{fig:thickfilm} and are again represented by dashed lines.
The expected positions for bcc Fe$_{100-x}$Ru$_{x}$ ($x=13$ and 17) are very close together, as well as the positions for hcp Fe$_{100-x}$Ru$_{x}$ ($x=17$, 20, and 23). Consequently, we only included the expected diffraction peak positions for bcc Fe$_{87}$Ru$_{13}$ and hcp Fe$_{77}$Ru$_{23}$ in \cref{fig:thinfilm}, which are the same as those plotted in \cref{fig:thickfilm}.
As expected, the lattice structure of the \SI{20}{nm}-thick films (\cref{fig:thinfilm}(a)) is the same as that of the \SI{100}{nm}-thick films (\cref{fig:thickfilm}(b)).
For both thicknesses, diffraction peaks corresponding only to the bcc lattice structure can be detected for Fe$_{87}$Ru$_{13}$, and peaks corresponding only to the hcp lattice structure can be detected for Fe$_{77}$Ru$_{23}$.
For the first intermediate composition in the \SI{20}{nm}-thick measurements, Fe$_{83}$Ru$_{17}$, both bcc and hcp reflections are visible, but for Fe$_{80}$Ru$_{20}$, only hcp reflections are clearly visible.

One of the most promising applications of FeRu takes advantage of the recently-discovered property that FeRu spacer layers can noncollinearly couple two ferromagnetic layers; this allows for significant improvement in the performance of spintronic devices such as spin-transfer-torque magnetic random-access memory \cite{nunn2020control}. For these devices to be compatible with the modern CMOS fabrication process, they need to be annealed at \SI{400}{\celsius}. Hence, in this paper, we explored the structural properties of \xFed[20] for $x=13$ and 20 before and after annealing at \SI{400}{\celsius} for 1~hour.
The in-plane $\theta$-$2\theta$ measurements shown in \cref{fig:thinfilm}(b) show that the as-deposited and annealed films have the same lattice structure. For Fe$_{87}$Ru$_{13}$, annealing causes a slight decrease in the bcc FeRu \{110\}, \{002\}, and \{112\} diffraction peak intensities along with the emergence of a trace of hcp phase, which could indicate that a small amount of the bcc phase transforms to the hcp phase as the temperature is increased. In the equilibrium phase diagram of FeRu \cite{Swartzendruber1983}, a mixture of the bcc and hcp phases is present for this Ru concentration at \SI{400}{\celsius}.
For Fe$_{80}$Ru$_{20}$, annealing causes an increase in the hcp FeRu \{100\}, \{110\}, and \{200\} diffraction peak intensities, which could indicate that annealing improves the texture of the hcp lattice structure of FeRu. A small amount of the bcc phase could be present in the as-deposited Fe$_{80}$Ru$_{20}$ film but not detectable by XRD, and as it transforms into the hcp phase during annealing, the hcp diffraction peak intensities increase.
These annealing results are encouraging, as they show that hcp FeRu layers annealed at temperatures of up to \SI{400}{\celsius} will retain their lattice structure in the composition range of interest for noncollinear coupling of Co$|$Fe$_{100-x}$Ru$_{x}|$Co films, $x\geq20$ \cite{nunn2020control}. This assumes that the diffusion of Co, Fe, and Ru across the Co$|$Fe$_{100-x}$Ru$_{x}$ interfaces is negligible at and below \SI{400}{\celsius}.

In the range from $x=0$ to 8, additional in-plane and out-of-plane $\theta$-$2\theta$ measurements were performed on as-deposited \xFed[100]. \Cref{fig:xraytrend} shows the $2\theta$ value of the bcc Fe or FeRu \{110\} diffraction peak measured for these films and the corresponding lattice constant as a function of $x$. 
Since Fe and Ru form a solid solution in this composition range and Fe atoms are smaller than Ru atoms, an increase in Ru concentration leads to an increase in lattice constant. By Bragg's law, this translates to a decrease in the angle $2\theta$ at which diffraction peaks are detected. As mentioned above, this effect is also visible in \cref{fig:thickfilm}(b) as the Ru concentration increases from 23 to \SI{50}{\at}.
Along with the experimental $2\theta$ and lattice constant values in \cref{fig:xraytrend} are two linear functions, both of which assume that \xFe{} remains bcc for $0\leq x\leq 8$. The bcc Ru interpolation, represented by the solid green line, is a linear function constructed from the following two points: the measured lattice constant of our Fe film obtained from the average of the in-plane and out-of-plane measurements ($x=0$) and the lattice constant of a hypothetical bcc unit cell of Ru atoms with radius \SI{.134}{\nano\meter} ($x=100$).
The Fe$_{96}$Ru$_{4}$ extrapolation, represented by the dashed yellow line, is a linear function constructed from the average of the in-plane and out-of-plane measurements of our Fe film ($x=0$) and our Fe$_{96}$Ru$_4$ film ($x=4$). We used these two points because the equilibrium phase diagram shows that up to \SI{4}{\at} of Ru can be added to \bFe{} at elevated temperatures before the hcp lattice structure begins to appear \cite{Swartzendruber1983}.
The measured lattice constant at $x=8$ is larger than that expected from both linear functions, indicating that Ru atoms are substituting Fe atoms in the bcc lattice up to at least \SI{8}{\at} of Ru in Fe.
This is not unexpected, as films deposited by sputtering are likely not in equilibrium. Furthermore, the out-of-plane and in-plane measurements in \cref{fig:xraytrend} yield similar results, indicating little difference in the lattice spacing for the two directions and suggesting that there is no distortion of the cubic structure.

	\subsection{Transmission electron microscopy}

\begin{figure*}[tbph]
			\centering \includegraphics[width=17.2cm]{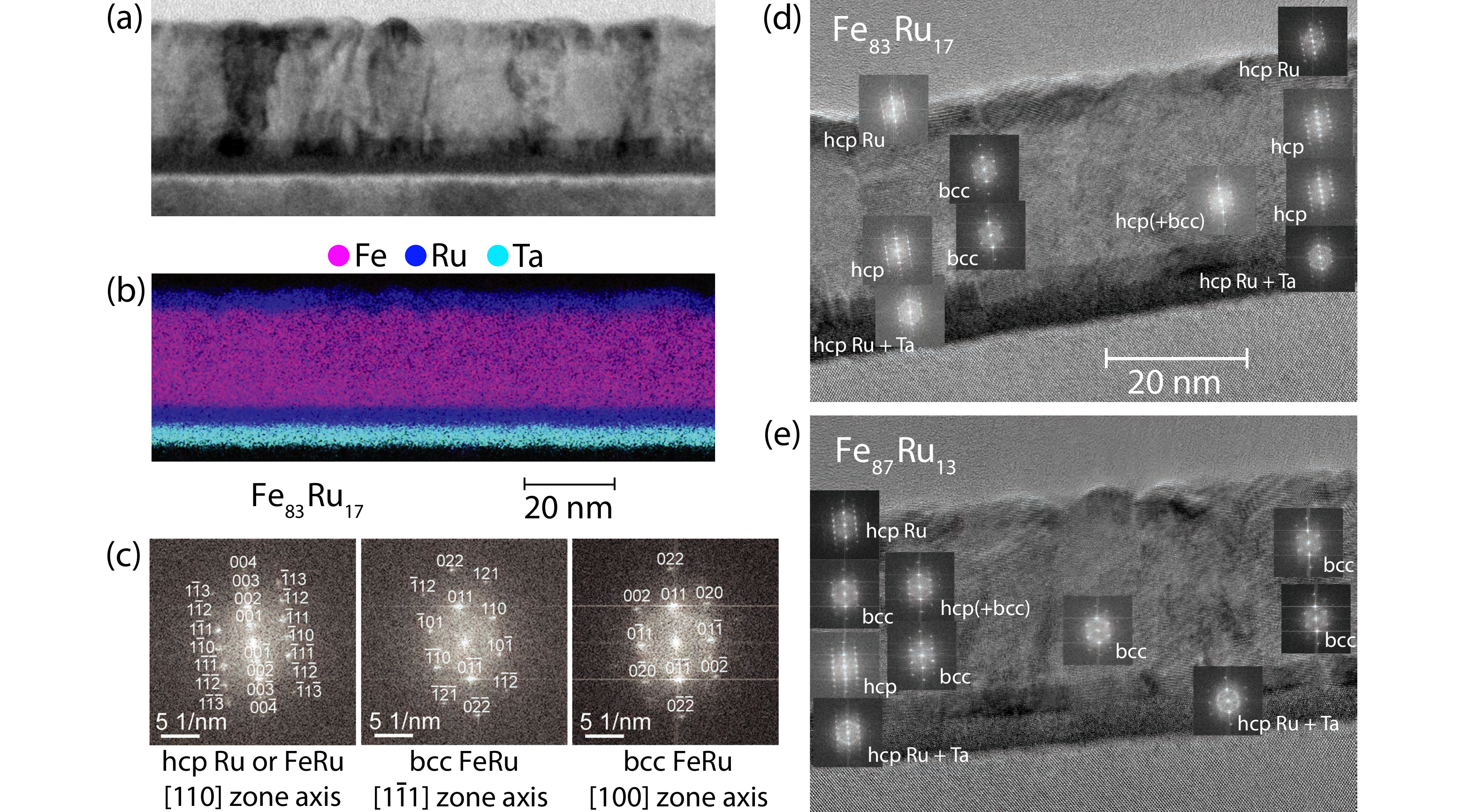}
            \caption{(a) Bright-field TEM and (b) STEM-EDXS-based element distribution images of Ta(3.5)$|$Ru(3.5)$|$Fe$_{83}$Ru$_{17}(20)|$Ru(3.5) with scale bar below, and (c) fast Fourier transform (FFT) patterns of hcp FeRu and Ru along the [110] zone axis and bcc FeRu along the [1$\bar{1}$1] and [100] zone axes.
            Example high-resolution TEM images of Ta(3.5)$|$Ru(3.5)$|$Fe$_{100-x}$Ru$_{x}(20)|$Ru(3.5) for (d) $x=17$ and (e) $x=13$ with scale bar in (d), including FFT patterns of various regions in the Ru and FeRu layers.}
 		\label{fig:TEM}
 	\end{figure*}
 	
Cross-sectional transmission electron microscopy (TEM) imaging was performed on \xFed[20] for $x=13$ and 17 to identify the distribution of the hcp and bcc phases in FeRu. Fe$_{83}$Ru$_{17}$ was selected because it consists of a mixture of hcp and bcc FeRu, as is evident from our X-ray diffraction (XRD) measurements in \cref{fig:thinfilm}(a).
Fe$_{87}$Ru$_{13}$ was selected because this alloy has the highest Ru concentration for which hcp diffraction peaks are not clearly detected in the XRD patterns.

\Cref{fig:TEM}(a) and (b) show a representative bright-field TEM image and the corresponding element distributions obtained by EDXS-based spectrum imaging analysis for \Fed{83}{17}{20}, respectively. The EDXS measurements confirm that the distribution of Fe and Ru in the FeRu film is homogeneous within measurement uncertainty. This suggests that the coexistence of the hcp and bcc phases in the FeRu films is not due to variation of the composition.

HRTEM images presented in \cref{fig:TEM}(d) and (e) for \Fed{83}{17}{20} and \Fed{87}{13}{20}, respectively, are further analyzed by FFT analysis.
The FFT patterns are placed on the HRTEM images at the locations at which they are calculated. Representative FFT patterns of hcp Ru and FeRu in [110] zone axis geometry and bcc FeRu in [1$\bar{1}$1] and [100] zone axis geometry are displayed separately in \cref{fig:TEM}(c).
In both films, the FFT patterns of Ru clearly show that Ru has hcp lattice structure, which is in agreement with the XRD measurements.
FFT analysis also reveals that both FeRu films consist of grains that have entirely hcp and entirely bcc crystal structure; furthermore, each film has indications of individual grains showing both phases. Since hcp diffraction peaks were not detectable in the XRD patterns of Fe$_{87}$Ru$_{13}$, it is evident that TEM analysis is more precise than XRD in detecting the presence of small amounts of a secondary phase.

We also performed cross-sectional TEM on Ta$|$Ru$|$Fe(100)$|$Ru to analyze the diffusion of Ru into the pure Fe film. The bright-field TEM and superimposed EDXS-based element distribution maps for a representative section of the sample are shown in the Supplemental Material \cite{supplemental}. 
Quantifying the EDXS signal from the whole Fe layer, a Ru content of \SI{0.1}{\at} was determined. This could either be due to the diffusion of Ru into the Fe layer during the sputtering process or due to the introduction of Ru into the Fe layer during the TEM lamella preparation process.

	\section{Magnetic Properties}

		\subsection{M\"ossbauer spectroscopy}
        \label{sec:moss}

\begin{table}[tbph]
\centering
\resizebox{0.48\textwidth}{!}{%
\begin{tabular}{cccccc}\toprule
\multirow{2}{*}{\textbf{Film}} & \textbf{IS} & \textbf{QS} & \boldmath{$B_\textbf{HF}$} & {\boldmath{\phiMS}} & {\textbf{PM}}\\
 & (\SI{}{\milli \meter / \second}) & (\SI{}{\milli \meter / \second}) & (\SI{}{\tesla}) & (\SI{}{\degree}) & (\%)\\ \midrule
§Fe$_{50}$Ru$_{50}$ & \,-0.047(2)& ~0.219(3) & & &100\\ \midrule
§Fe$_{77}$Ru$_{23}$& \,-0.074(1)& ~0.188(2) & & &100\\ \midrule
§Fe$_{80.5}$Ru$_{19.5}$ & \,-0.07(1)~\, & 0.14(2)\,  &&&100 \\ \midrule
  \begin{tabular}{@{}c@{}}
    §Fe$_{80.5}$Ru$_{19.5}$* \\
    (at \SI{9.9}{K}) \\
  \end{tabular} & \,~0.044(2) & 0.31(1)\,  &&&100 \\ \midrule
\multirow{2}{*}[-0.4065ex]{§Fe$_{81.5}$Ru$_{18.5}$} & \,~0.028(1)&& \,30.2(1)~\, & 40(2)&\multirow{2}{*}{61(2)}\\ \cmidrule(lr){2-5}
& \,-0.072(1)& ~0.171(3) && \\ \midrule
\multirow{2}{*}[-0.4065ex]{
  \begin{tabular}{@{}c@{}}
    §Fe$_{81.5}$Ru$_{18.5}$* \\
    (at \SI{9.7}{K}) \\
  \end{tabular}} 
 & 0.19(1)&& \,32.3(6)~\, & 35(2) &\multirow{2}{*}{65(1)}\\ \cmidrule(lr){2-5}
& \,-0.045(2)& 0.30(1)\, && \\ \midrule
\multirow{2}{*}[-0.4065ex]{§Fe$_{87}$Ru$_{13}$} & \,~0.037(4)&  & \,32.2(1)~\, & 45(1)&\multirow{2}{*}{4(1)}\\ \cmidrule(lr){2-5}
 & -0.37(2)\,\,& 0.79(4)\, & & \\ \midrule
§Fe$_{92}$Ru$_{8}$* & \,~0.027(3)&& \,31.5(1)~\, & 44(1) &3.0(5)\\ \midrule
  \begin{tabular}{@{}c@{}}
    §Fe$_{92}$Ru$_{8}$* \\
    (at \SI{0.166}{T}) \\
  \end{tabular} & \,~0.032(3)&& \,31.2(3)~\,& 47(1)&3.0(5)\\ \midrule
§Fe$_{96}$Ru$_4$* & \,~0.019(3)&& \,31.9(1)~\, & 51(1)&2.5(7)\\ \midrule
  \begin{tabular}{@{}c@{}}
    §Fe$_{96}$Ru$_{4}$* \\
    (at \SI{0.166}{T}) \\
  \end{tabular} & \,~0.018(5)&& \,32.9(3)~\,& 55(1)&2(1)\\ \midrule
§Fe*& \,~0.000(4)&& \,32.83(3)& 68(2)&0\\ \midrule
  \begin{tabular}{@{}c@{}}
    §Fe* \\
    (at \SI{0.166}{T}) \\
  \end{tabular} & \,~0.003(5)&& \,32.61(4)& 67(2)&0\\ \midrule 
Ta$|$Fe$|$Ta& \,~0.007(1)&& \,33.39(2)& 73(1)&0\\ \midrule
Fe & \,-0.003(1)&& \,33.42(1)& 80(1)&0\\ \midrule
Fe* & \,-0.002(1)&& \,33.13(1)& 82(1)&0\\ \midrule
  \begin{tabular}{@{}c@{}}
    Fe* \\
    (at \SI{0.166}{T}) \\
  \end{tabular} & \,-0.001(1) && \,32.38(1)& 85(2)&0\\ \bottomrule
\end{tabular}}
\caption{Isomer shift (IS), quadrupole splitting (QS), hyperfine magnetic field ($B_\text{HF}$), angle between average magnetization and film normal (\phiMS), and amount of paramagnetic phase (PM) were extracted from M\"ossbauer spectroscopy measurements of the films in the left column. Films marked with § signify that the full structure is Ta(3.5)$|$Ru(3.5)$|$Fe$_{100-x}$Ru$_{x}$(100)$|$Ru(3.5). Rows marked with * indicate that the sample was measured via transmission M\"ossbauer spectroscopy, and rows without * were measured via conversion electron M\"ossbauer spectroscopy. Unless specified, each measurement is performed at zero external magnetic field.
}
\label{tab:moss}
\end{table}

M\"ossbauer measurements were used to study \xFed[100]{} films, with transmission M\"ossbauer spectroscopy (TMS) performed on samples with $x=0$, 4, and 8 at room temperature and conversion electron M\"ossbauer spectroscopy (CEMS) performed on samples with $x=13$, 18.5, 19.5, 23, and 50 at room temperature.
Furthermore, CEMS was performed on Ta$|$Fe(100)$|$Ta, and CEMS and TMS were both performed on a single Fe(100) layer, all at room temperature.
Additional TMS measurements of the single Fe(100) layer and of the \xFed[100]{} samples with $x=0$, 4, and 8 were performed in the presence of a \SI{0.166}{\tesla} in-plane magnetic field, and TMS measurements of the samples with $x=18.5$ and 19.5 were performed at 9.7 and \SI{9.9}{K}, respectively.

For all measured films, the M\"ossbauer spectra were fitted to determine the isomer shift (IS), quadrupole splitting (QS), and hyperfine magnetic field ($B_\text{HF}$). Paramagnetic components of M\"ossbauer spectra have a $B_\text{HF}$ of zero, and films with a cubic lattice structure have a QS of zero.
The angle between the gamma ray (perpendicular to the film plane) and the average magnetization of the film, \phiMS{}, was determined by comparing the peak intensities of the second and fifth lines with the other lines in the spectra.
The intensities of the second and fifth lines depend only on the angle between the gamma ray (perpendicular to the film plane) and hyperfine field and are not affected by the in-plane orientation of the hyperfine field or the sign of the component of the hyperfine field along the gamma ray (parallel and antiparallel are equivalent). As a result, M\"ossbauer measurements yield the same angle if the out-of-plane magnetization components of the film are all pointing in the same direction as the gamma ray, all pointing in the opposite direction of the gamma ray, or any combination of the two.
In contrast, VSM measurements with the external magnetic field applied perpendicular to the film normal treat these cases differently. In VSM measurements, the projections onto the field direction of all the magnetization components are summed, taking into account that projections in the same direction as the external magnetic field are positive and projections in the opposite direction are negative (see Sec.~\ref{sec:MH}).
The four above-described parameters extracted from M\"ossbauer measurements\textemdash IS, QS, $B_\text{HF}$, and \phiMS{}\textemdash as well as the amount of paramagnetic phase in the film (PM), are tabulated in \cref{tab:moss} for all films measured by M\"ossbauer spectroscopy.

  \begin{figure}[tbph]
			\centering
			\includegraphics[width=8.6cm]{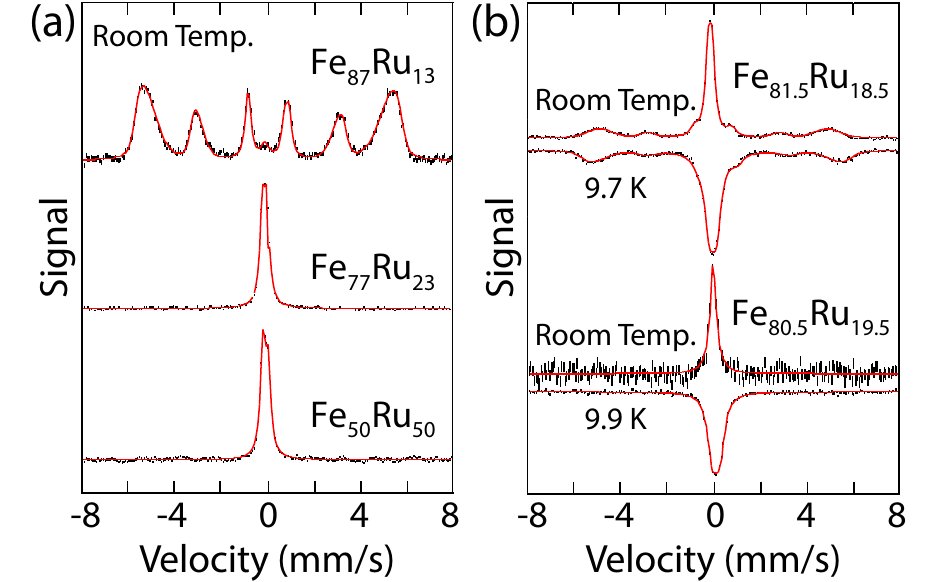}
            \caption{(a) Conversion electron M\"ossbauer spectroscopy (CEMS) measurements at room temperature of Ta(3.5)$|$Ru(3.5)$|$Fe$_{100-x}$Ru$_{x}$(100)$|$Ru(3.5) for $x=13$, 23, and 50 and (b) CEMS measurements at room temperature and transmission M\"ossbauer spectroscopy (TMS) measurements at low temperature of Ta(3.5)$|$Ru(3.5)$|$Fe$_{100-x}$Ru$_{x}$(100)$|$Ru(3.5) for $x=18.5$ and 19.5. All measurements are performed at zero external magnetic field.}
 		\label{fig:moss1}
 	\end{figure}

\Cref{tab:moss} shows that \xFed[100]{} is fully paramagnetic for $x\geq19.5$, as indicated by PM being \SI{100}{\%} and $B_\text{HF}$ being absent. This can be visualized in the CEMS measurements in \cref{fig:moss1}, which show one central peak for $x=19.5$, 23, and 50.
When compared with the X-ray diffraction (XRD) measurements, it follows that \hFeRu{} is paramagnetic at room temperature.
When the Ru concentration decreases from 19.5 to \SI{18.5}{\at}, $B_\text{HF}$ appears and the amount of paramagnetic phase in the film drops from 100 to \SI{61}{\%}. The mixture of the paramagnetic and magnetic phase in Fe$_{81.5}$Ru$_{18.5}$ can also be seen in the CEMS measurements in \cref{fig:moss1}(b).
At this concentration, $B_\text{HF}$ is \SI{30.2\pm0.1}{T}, which is already about $90\%$ of the hyperfine field known for pure Fe, \SI{33}{T} \cite{violet2003mossbauer}.
As the Ru concentration in \xFe{} decreases further from $18.5$ to \SI{13}{\at}, $B_\text{HF}$ increases to \SI{32.2\pm0.1}{T} and only \SI{4}{\%} of the film is paramagnetic.
The small amount of paramagnetic phase remaining in the film is indicated by the small central peak in the CEMS measurements of Fe$_{87}$Ru$_{13}$ in \cref{fig:moss1}(a).
Though the presence of the hcp phase in Fe$_{87}$Ru$_{13}$ was not detected by XRD, it was detected by the more precise method, TEM, confirming that the paramagnetic contribution observed by M\"ossbauer spectroscopy is due to the \hFeRu{} phase. M\"ossbauer spectroscopy also shows that films for which the magnetic phase is detected, i.e., \xFe{} with $x\leq18.5$, also have an average magnetization that is tilted out of the film plane.
The average magnetization angle from the film normal, \phiMS{}, does not change significantly until the Ru concentration is less than \SI{8}{\at}.
For Fe$_{81.5}$Ru$_{18.5}$, CEMS measurements give an $\phiMS=\SI{40\pm2}{\degree}$ from the film normal.
For Fe$_{87}$Ru$_{13}$, CEMS measurements yield $\phiMS=\SI{45 \pm1}{\degree}$, and for Fe$_{92}$Ru$_{8}$, TMS measurements give $\phiMS=\SI{44 \pm1}{\degree}$.

When Fe$_{81.5}$Ru$_{18.5}$ is cooled to \SI{9.7}{K}, $B_\text{HF}$ of the magnetic phase increases from \SI{30.2\pm0.1}{T} to \SI{32.3\pm0.6}{T}, but the amounts of the paramagnetic and magnetic phases remain practically the same (see \cref{tab:moss} and \cref{fig:moss1}(b)). Additionally, \cref{tab:moss} and \cref{fig:moss1}(b) show that Fe$_{80.5}$Ru$_{19.5}$ remains fully paramagnetic when cooled to \SI{9.9}{K}.
This indicates that Fe$_{100-x}$Ru$_{x}$ is paramagnetic above \SI{9.9}{K} for $x\geq19.5$, which differs from previous neutron diffraction studies that reported an incommensurate antiferromagnetic state of Fe$_{71}$Ru$_{29}$ with a N\'eel temperature of \SI{124}{K} \cite{PetrilloND2018}.

\begin{figure}[tbph]
			\centering
        \includegraphics[width=8.6cm]{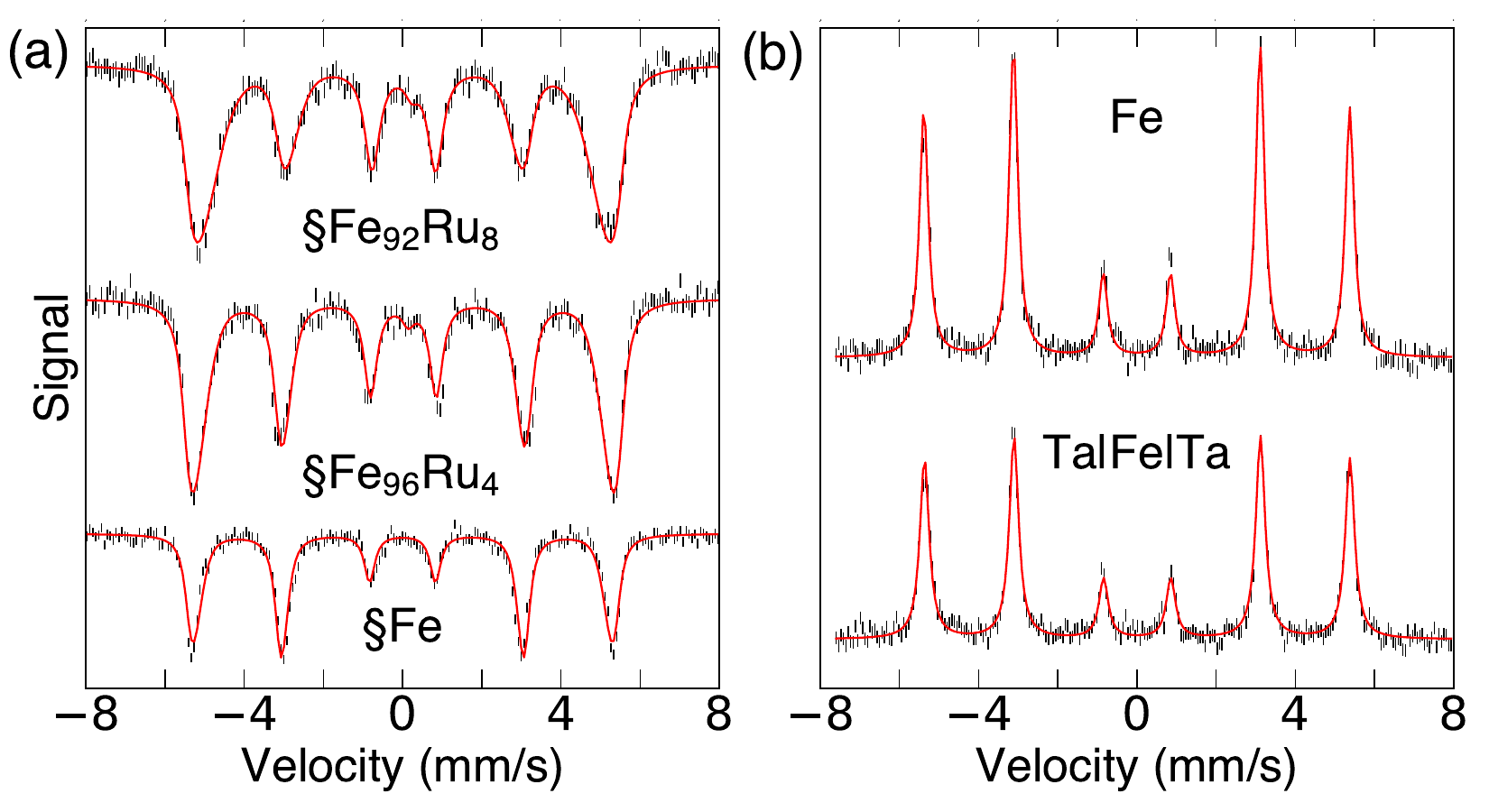}
             \caption{(a) Transmission M\"ossbauer spectroscopy (TMS) measurements at room temperature of Ta(3.5)$|$Ru(3.5)$|$Fe$_{100-x}$Ru$_{x}$(100)$|$Ru(3.5) for $x=0$, 4, and 8, and (b) conversion electron M\"ossbauer spectroscopy (CEMS) at room temperature of Fe(100) (no surrounding layers) and Ta(3.5)$|$Fe(100)$|$Ta(3.5).  All measurements are performed at zero external magnetic field.}
 		\label{fig:moss2}
 	\end{figure}

The TMS measurements of \xFed[100]{} for $x=4$ and $8$ in \cref{fig:moss2}(a) show that these films have only a trace of the paramagnetic phase, which disappears for $x=0$. This is confirmed by the amount of paramagnetic phase tabulated in \cref{tab:moss}.
Comparing these results with the XRD measurements, we can confirm that bcc Fe and FeRu are magnetic. 
Since the magnetic phase has cubic structure (bcc), quadrupole splitting is zero for $x\leq8$.
\Cref{tab:moss} shows that the hyperfine field increases as $x$ decreases from 8 to \SI{0}{\at} of Ru in Fe and reaches approximately \SI{33}{T}, the hyperfine field previously measured for pure Fe \cite{violet2003mossbauer}.
In the same Ru concentration range, the average magnetization direction moves towards the plane, as seen in \cref{tab:moss}. For $x=8$, TMS measurements give an angle of $\phiMS=\SI{44 \pm1}{\degree}$ from the film normal. For $x=4$, \phiMS{} increases to \SI{51 \pm1}{\degree}, and for $x=0$, \phiMS{} increases further to \SI{68 \pm2}{\degree} from the film normal.
This result for Ta$|$Ru$|$Fe(100)$|$Ru is very unexpected. Due to the relatively small magnetocrystalline anisotropy of \bFe{} as compared to the large in-plane shape anisotropy of the studied Fe films, the magnetization is expected to lie in the plane. The large shape anisotropy arises from the large $M_s$ of the studied Fe films.
One possible explanation of the out-of-plane magnetization components measured in Ta$|$Ru$|$Fe(100)$|$Ru is the presence of a very small amount of Ru in the Fe film. Quantification of the STEM-EDXS analysis of this film (Supplemental Material \cite{supplemental}) indicated that the Fe layer may contain up to \SI{0.1}{\at} of Ru.

To eliminate any contribution from Ru in the Fe film, CEMS was performed on an \SI{100}{nm}-thick Fe film surrounded only by \SI{3.5}{nm}-thick Ta, and CEMS and TMS were both performed on a single \SI{100}{nm}-thick Fe layer. The parameters of these three measurements are presented in \cref{tab:moss}, and the CEMS measurements of both films are displayed in \cref{fig:moss2}(b). 
The hyperfine fields for both films are approximately \SI{33}{T}, in agreement with the hyperfine field known for Fe \cite{violet2003mossbauer}.
However, the average magnetization direction as compared to that of Ta$|$Ru$|$Fe$|$Ru moves closer to the plane for Ta$|$Fe$|$Ta ($\phiMS=\SI{73\pm1}{\degree}$ from the normal), and closer still for the single Fe layer ($\phiMS=\SI{80\pm1}{\degree}$ from CEMS and \SI{82\pm1}{\degree} from TMS).
Since domain walls have out-of-plane magnetization components that could be responsible for these observations, an in-plane magnetic field of \SI{0.166}{T} was applied to the single Fe layer to expel all domains and TMS was again performed. As a result of the application of the magnetic field, \phiMS{} of the single Fe layer only increased to \SI{85\pm2}{\degree} from the film normal, showing that the presence of domain walls only changes the average magnetization direction by a few degrees.
For confirmation, we also measured \xFed[100]{} for $x=0$, 4, and 8 with TMS in the presence of the \SI{0.166}{T} in-plane magnetic field and, as with the single Fe layer, the expelling of magnetic domains only changed \phiMS{} by a few degrees.
Micromagnetic simulations of an \SI{100}{nm}-thick Fe layer grown along the $\langle110\rangle$ directions were performed to determine whether the magnetization has an out-of-plane component and are presented in the Supplemental Material \cite{supplemental}.
In the absence of an external magnetic field, the simulations yielded an average magnetization of \SI{84\pm10}{\degree} from the film normal. However, when a \SI{0.05}{T} in-plane magnetic field was applied in the simulations, magnetic domains were expelled, and the film was fully saturated along the field direction with no out-of-plane magnetization component.

\subsection{Vibrating sample magnetometry}
\label{sec:MH}

\begin{figure*}[tbph]
			\centering
			\includegraphics[width=17.2cm]{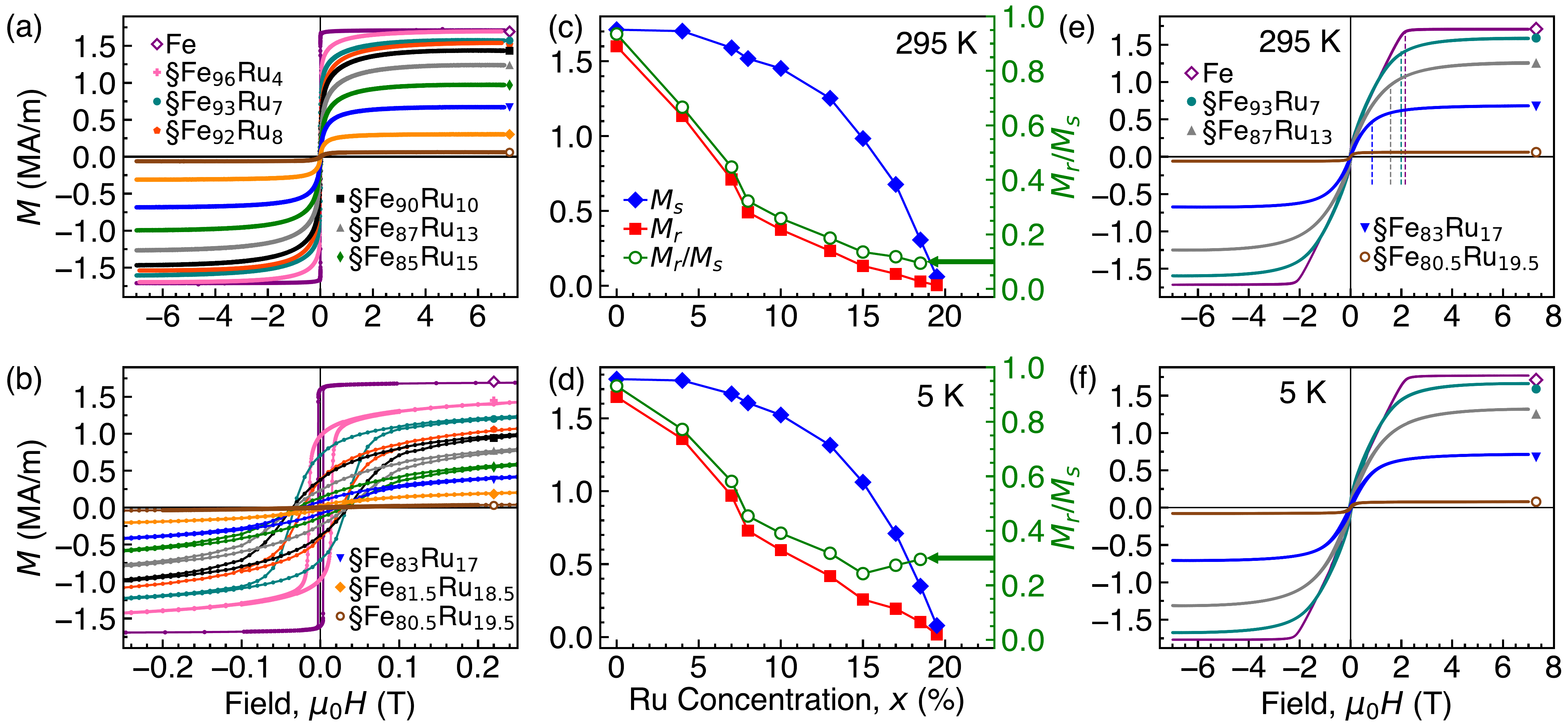}
            \caption{$M(H)$ measurements with magnetic field applied parallel to the film surface over (a) the full measured field range and (b) a zoomed-in field range to highlight the hysteresis loops of as-deposited Fe(100), §\xFe{}(100) for $x=4$ and 8, and §\xFe{}(20) for $x= 7$, 10, 13, 15, 17, 18.5, and 19.5, where § signifies that the full structure is Ta(3.5)$|$Ru(3.5)$|$Fe$_{100-x}$Ru$_x(d)|$Ru(3.5). $M_s$, $M_r$, and $M_r/M_s$ from the $M(H)$ measurements in (a,b) are plotted at (c) 295 and (d) \SI{5}{K}.
            $M(H)$ measurements with magnetic field applied perpendicular to the film surface of as-deposited Fe(100) and §\xFe{}(20) for $x=7$, 13, 17, and 19.5 at (e) 295 and (f) \SI{5}{K}. The dashed lines in (e) represent the calculated demagnetizing fields for Fe(100) and §\xFe{}(20) ($7\leq x\leq17$) films. }
 		\label{fig:MHcurves}
 	\end{figure*}

Vibrating sample magnetometry (VSM) measurements were used to determine the magnetization of \xFed[d], Ta$|$Fe(100)$|$Ta, and Fe(100) films as a function of an external magnetic field applied either parallel or perpendicular to the film surface and at a temperature of 5 or \SI{295}{K}. From the measurements, we obtained the saturation magnetization, $M_s$, and the remanent magnetization along the field direction, $M_r$. The remanent-to-saturation magnetization ratio, $M_r/M_s$, allows us to infer information about the magnetic structure of the \xFe{} films.
Additionally, for selected structures, we found the external magnetic field above which the magnetization is reversible, $H_\text{rev}$. Above this field, domains are expelled from the film and do not contribute to the $M(H)$ measurements.

\Cref{fig:MHcurves}(a,b) shows the $M(H)$ measurements at \SI{295}{K} of Fe(100), \xFed[100] for $x=4$ and 8, and \xFed[20] for $x =$ 7, 10, 13, 15, 17, 18.5, and 19.5, with the magnetic field applied parallel to the film surface (referred to as in-plane). $M_r$, $M_s$, and the $M_r/M_s$ ratio for each film are plotted as a function of Ru concentration in \cref{fig:MHcurves}(c) and (d) at 295 and \SI{5}{K}, respectively.
When a very large in-plane magnetic field is applied, the magnetic moments lie along the field direction, producing a net magnetization of $M_s$ in the field direction. As the field is reduced to zero, the magnetic moments may deviate from the field direction, resulting in a remanent magnetization along the field direction of $M_r\leq M_s$. Therefore, the ratio $M_r/M_s$ provides information on the distribution of magnetic moments in the film in the absence of an external magnetic field. $M_r/M_s=1$ means that the magnetic moments are fully aligned along the field direction after the field is removed, and $M_r/M_s=0$ means that there is no net magnetization along the field direction after the field is removed.
The observed $M_r/M_s$ values of the in-plane $M(H)$ measurements will be discussed in Sec.~\ref{sec:exp_res}.
 
It can be seen in \cref{fig:MHcurves}(c) that $M_s$ decreases with Ru concentration from approximately \SI{1710}{kA/m} for the single Fe layer to \SI{61}{kA/m} for Fe$_{80.5}$Ru$_{19.5}$ at room temperature.
Over the $0\leq x\leq10$ composition range of \xFe{}, the slow decrease in $M_s$ is due to Ru atoms substituting Fe atoms in the host bcc lattice. This agrees with the small decrease in the hyperfine field measured by M\"ossbauer spectroscopy over this composition range (\cref{tab:moss}).
The sharp decrease in $M_s$ observed in \xFe{} for $13\leq x\leq19.5$ is mainly due to the increasing amount of paramagnetic hcp FeRu and decreasing amount of magnetic bcc FeRu in the films. This is corroborated by our M\"ossbauer measurements in this composition range (\cref{tab:moss}), which show a large increase in the amount of paramagnetic phase in the film (from $4\%$ to $100\%$) and only a small change in the hyperfine field.

The $M_s$ measurements at \SI{5}{K} in \cref{fig:MHcurves}(d) are slightly higher than those at \SI{295}{K} for all plotted Ru concentrations, as is expected for ferromagnetic materials \cite{coey2010magnetism}.
\cref{fig:MHcurves}(d) also shows that if the ambient temperature is decreased from 295 to \SI{5}{K}, $M_r$ stays practically the same for the single Fe layer but increases for all FeRu alloys. Consequently, the $M_r/M_s$ ratio also increases as the ambient temperature is reduced to 5 K. The increase is most significant for films with the highest Ru concentrations.

\begin{figure}[tbph]
			\centering
            \includegraphics[width=8.6cm]{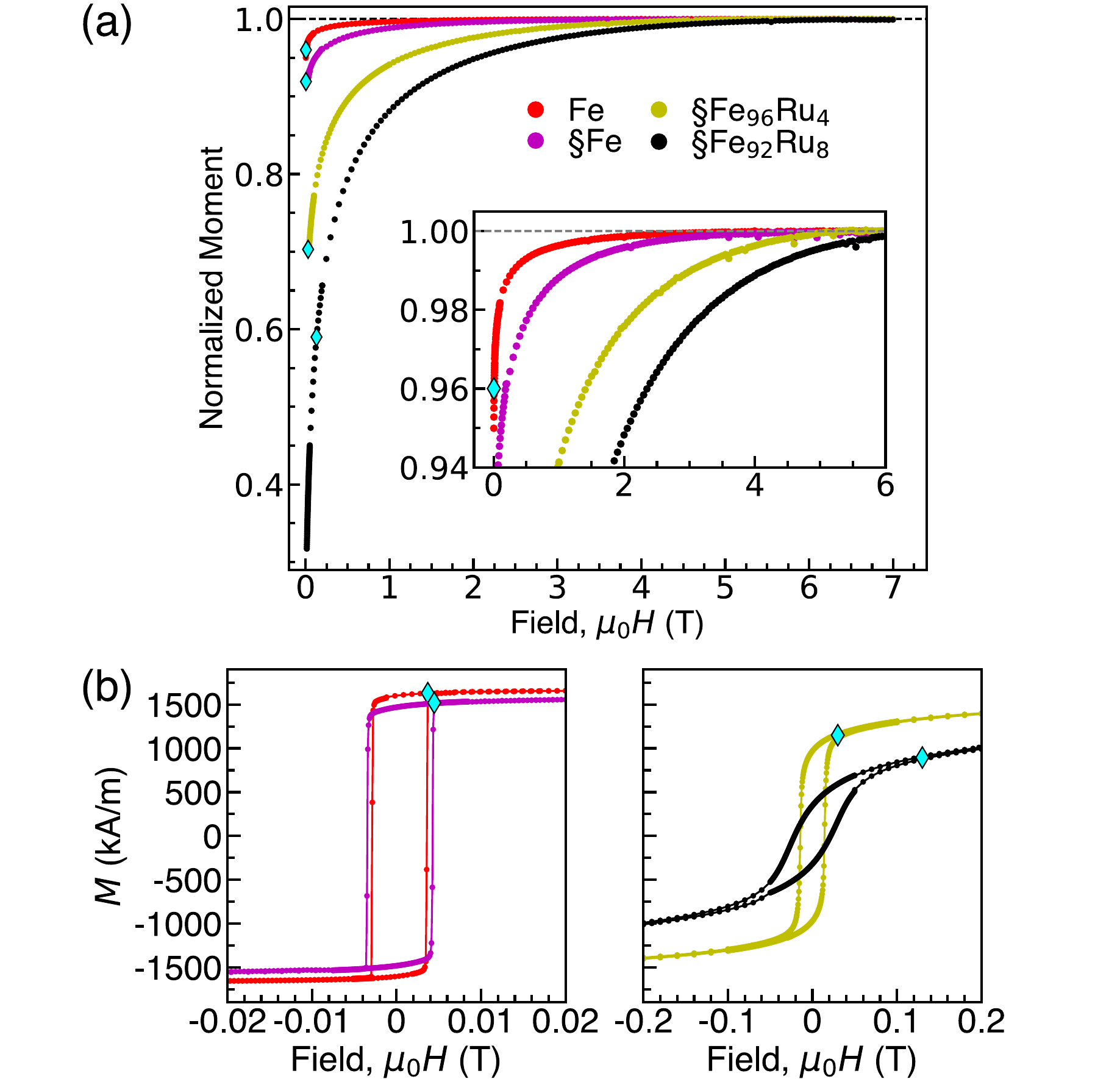}
            \caption{(a) $M(H)/M_s$ and (b) $M(H)$ measurements of as-deposited Fe(100) and of Ta(3.5)$|$Ru(3.5)$|$Fe$_{100-x}$Ru$_x$(100)$|$Ru(3.5) for $x=0$, 4, and 8, where the § in the legend signifies that the layer is grown on top of Ta$|$Ru and capped with Ru. The magnified inset plot in (a) highlights the variation in saturation field. The magnetic field ranges in (b) are chosen to highlight the hysteresis loops of the samples. Cyan diamonds are placed at the lowest field for which the hysteresis loop is reversible, indicating the removal of magnetic domains.}
 		\label{fig:saturationloop}
 	\end{figure}

From \cref{fig:MHcurves}(a,b), it is evident that the field required to saturate the FeRu alloys is much higher than that required to saturate the pure Fe film.
Large saturation fields are observed even for very low concentrations of Ru in Fe. To highlight the approach to saturation, \cref{fig:saturationloop}(a) shows the normalized $M(H)$ curves for $H\geq0$ of Fe(100) and of \xFed[100]{} for $x = 0$, 4, and 8. Even for Ta$|$Ru$|$Fe(100)$|$Ru, the field required to saturate the magnetization is significantly larger than that required for the single Fe layer with no surrounding layers. This could be due to Ru diffusing into the Fe layer, which is suggested by the TEM-based analysis in the Supplemental Material \cite{supplemental}.

Marked on each $M(H)$ curve in \cref{fig:saturationloop}(a) with a cyan diamond is the point at which the magnetization becomes reversible with the magnetic field, $M(H_\text{rev})$, meaning that all the magnetic domains have been expelled from the film.
These points are also indicated in \cref{fig:saturationloop}(b), which shows the $M(H)$ loops of the same films in \cref{fig:saturationloop}(a).
As the Ru concentration in the film increases, the field required to expel the domains increases; however, in each case, the domains are expelled in fields lower than \SI{0.13}{\tesla}. 
Magneto-optical microscopy measurements, which will be presented in Sec.~\ref{sec:opt}, show that domains disappear at \SI{\sim0.030}{\tesla} in Fe$_{96}$Ru$_4$, confirming that there are no domains above the $H_\text{rev}$ fields marked in \cref{fig:saturationloop}.
Since domains are not present in the reversible region, the presence of domains cannot explain the large saturation fields seen in \cref{fig:saturationloop}(a). 

\Cref{fig:MHcurves}(e) shows the $M(H)$ measurements of as-deposited Fe(100) and \xFed[20] for $x=7$, 13, 17, and 19.5 at \SI{295}{K}, with the magnetic field applied perpendicular to the film plane (referred to as perpendicular-to-plane).
When a large magnetic field is applied, the magnetic moments are saturated along the field direction.
When the field is removed, Fe and all FeRu alloys have an $M_r/M_s$ ratio of nearly zero, indicating that the net projection of magnetic moments along the field direction is nearly zero.
Combining this information with the out-of-plane magnetization detected by M\"ossbauer spectroscopy, we conclude that there are out-of-plane magnetization components pointing both up and down along the film normal that almost fully cancel each other out.

The field required to saturate the magnetization of the single Fe layer perpendicular to the film plane is slightly larger than the shape anisotropy or demagnetizing field, $\mu_0 M_s = \SI{2.15}{T}$, which is represented by the respective dashed line (purple) in \cref{fig:MHcurves}(e). In addition to shape anisotropy, Fe films have a much smaller magnetocrystalline anisotropy which is oriented along the $\langle100\rangle$ directions. The magnetocrystalline anisotropy field also contributes to the field required to saturate the sample.
When 7 to \SI{17}{\at} of Ru is added to the Fe film, the saturation fields for the perpendicular-to-plane measurements increase to above \SI{6}{T}, which is several times larger than the demagnetizing fields of the films (represented by additional dashed lines in \cref{fig:MHcurves}(e)).
The perpendicular-to-plane $M(H)$ measurements performed at \SI{5}{K}, presented in \cref{fig:MHcurves}(f), are very similar to those performed at \SI{295}{K}.

    \subsection{Magneto-optical Kerr effect microscopy}
\label{sec:opt}

\begin{figure}[tbph]
			\centering
		\includegraphics[width=8.6cm]{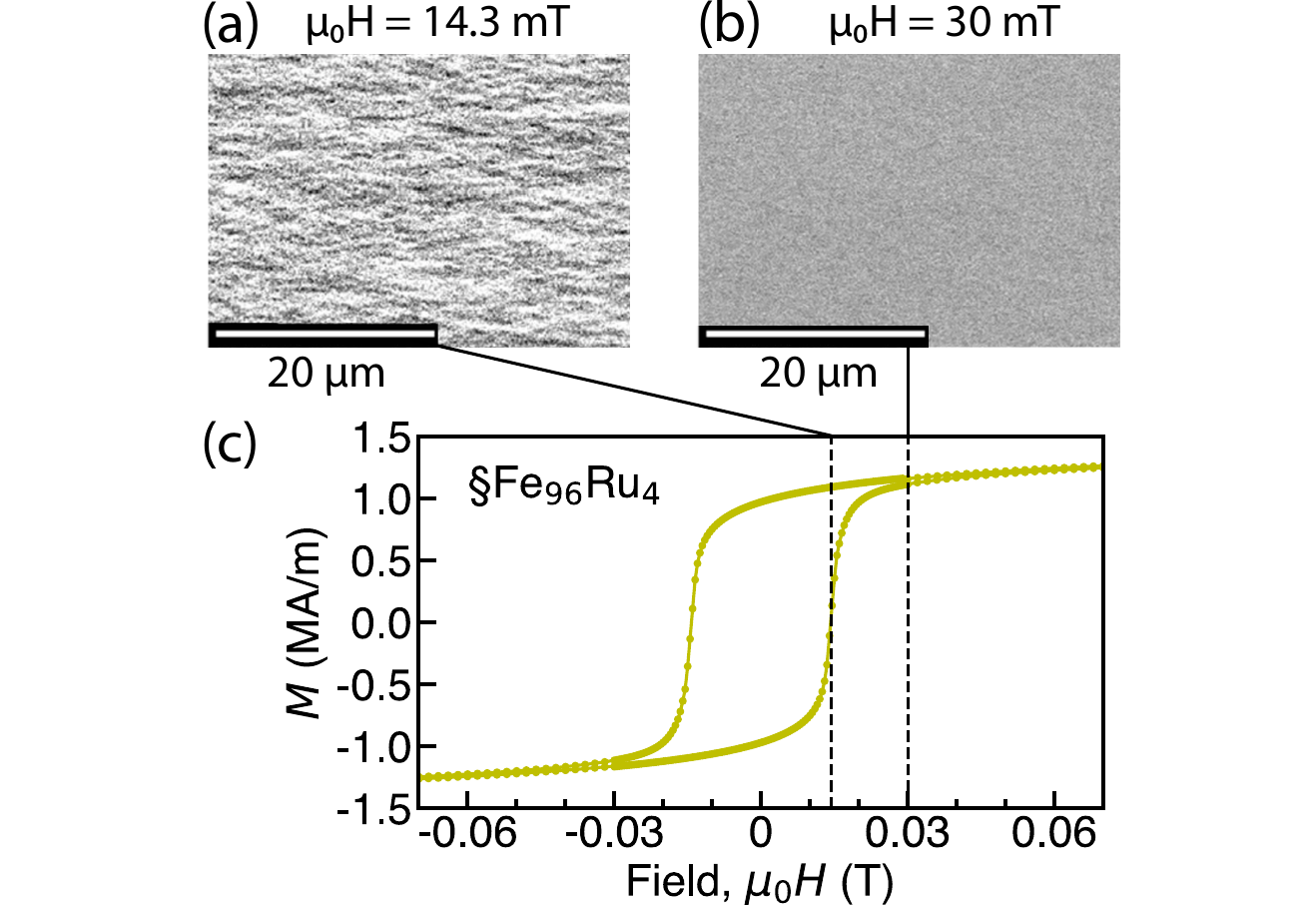}
            \caption{MOKE microscopy image of the Ta$|$Ru$|$Fe$_{96}$Ru$_{4}(100)$$|$Ru film surface with (a) \SI{14.3}{\milli \tesla} and (b) \SI{30}{\milli \tesla} applied parallel to the film surface. Small domains that are elongated along the film direction are visible at \SI{14.3}{\milli \tesla}, and no domains are visible as \SI{30}{\milli \tesla}.
            (c) $M(H)$ loop of Ta$|$Ru$|$Fe$_{96}$Ru$_{4}(100)$$|$Ru, which is also plotted in \cref{fig:saturationloop}(b).
            Dashed lines are placed at the field values that correspond with (a) and (b).
            }
 		\label{fig:MOKE}
 	\end{figure}

Magneto-optical Kerr effect (MOKE) microscopy measurements were performed on Ta$|$Ru$|$Fe$_{96}$Ru$_{4}(100)$$|$Ru to observe how the domain structure of the film evolves in an in-plane applied magnetic field.
Initially, a magnetic field of \SI{-1.1}{\tesla} was applied in the film plane, saturating the magnetic moments along the field direction. The field was subsequently removed and then swept up to \SI{1.1}{\tesla} while the domain structure was observed. 
\cref{fig:MOKE}(a) presents the MOKE microscopy image of Ta$|$Ru$|$Fe$_{96}$Ru$_{4}(100)$$|$Ru at an applied field of \SI{14.3}{\milli\tesla}. For reference, the $M(H)$ loop of the film, \cref{fig:MOKE}(c), is placed below the image. At \SI{14.3}{\milli\tesla}, the coercive field of the film, small domains textured along the field direction were observed. When the magnetic field reached \SI{30}{\milli\tesla}, the domains were entirely expelled, as can be seen in \cref{fig:MOKE}(b).
This field coincides with the point marked by a diamond in \cref{fig:saturationloop} at which the $M(H)$ loop of the Fe$_{96}$Ru$_{4}$ sample becomes reversible. Therefore, the reversible parts of our $M(H)$ loops are not affected by any out-of-plane magnetization components or noncollinearity in domain walls.
The texture of the domains follows the direction of the applied magnetic field, suggesting that the film is isotropic within the plane, i.e., there is no preferred magnetic anisotropy direction in the plane. The observed pattern is neither purely stripe nor ripple domain structure, but rather a domain structure typical for polycrystalline magnetic films that have magnetocrystalline anisotropy averaged out in the film plane.

\subsection{Lorentz transmission electron microscopy}

  \begin{figure}[tbph]
			\centering
		\includegraphics[width=8.6cm]{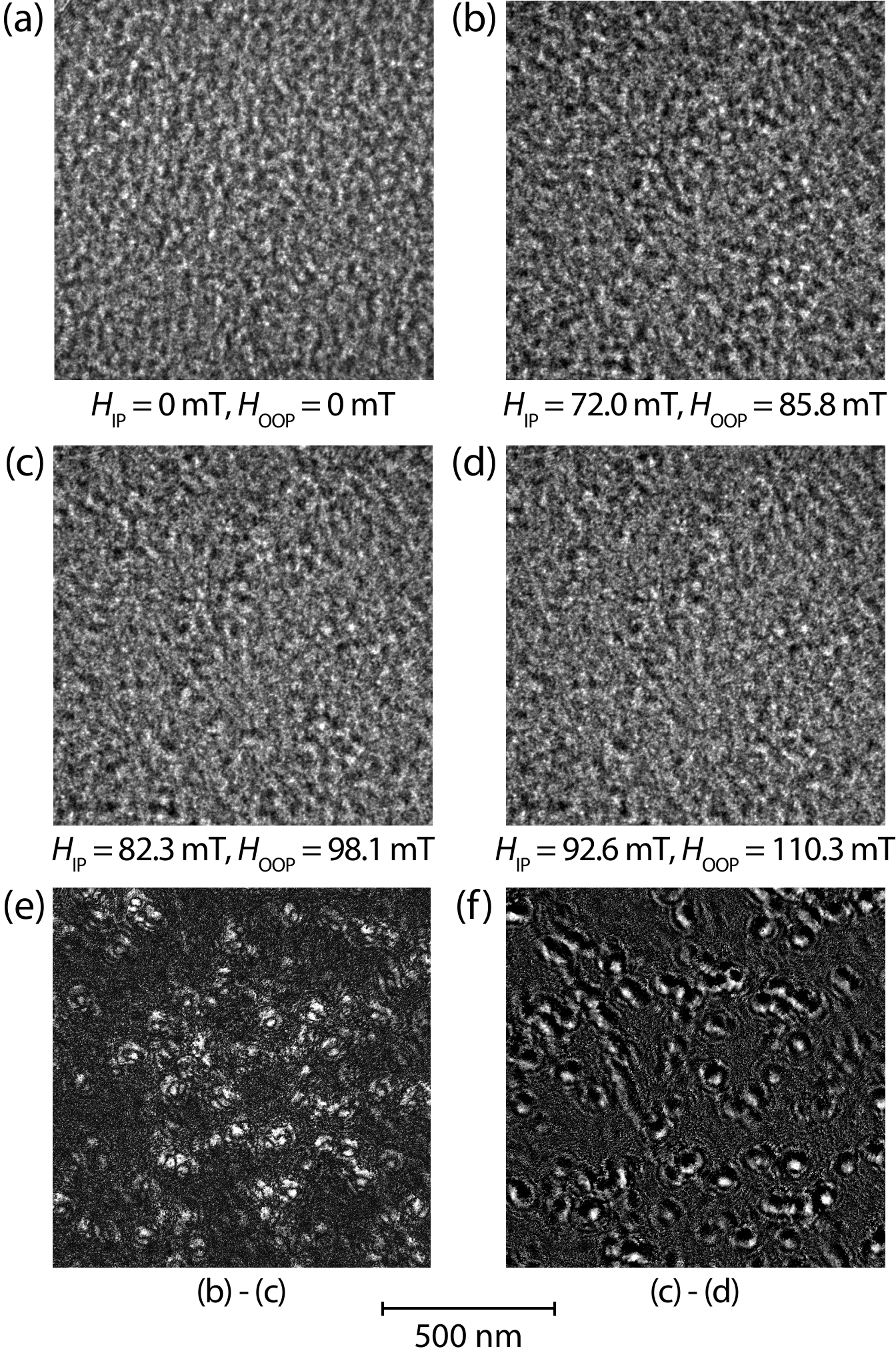}
            \caption{(a\textendash d) Lorentz transmission electron microscopy images taken at the same location on a Ta(3.5)$|$Ru(3.5)$|$Fe$_{92}$Ru$_{8}$(20)$|$Ru(3.5) film, for increasing applied magnetic field values, as labelled. (e,f) Corresponding contrast differences showing all the switching events that occurred due to the respective field increase. The scale bar at the bottom applies to all images. All images are taken at room temperature with a defocus of \SI{-1}{mm} and a tilting angle of $+\SI{40}{\degree}$.
            }
 		\label{fig:LTEM}
 	\end{figure}

The magnetization reversal process of a Ta$|$Ru$|$Fe$_{92}$Ru$_{8}$(20)$|$Ru film was investigated in more detail by Lorentz transmission electron microscopy (LTEM), which is able to probe the in-plane magnetization orientation at the nanoscale. The sample was tilted by a tilt angle of $\theta$ with respect to the film normal, yielding an in-plane magnetic field of $\HIP=\Happl\sin{\theta}$ and an out-of-plane magnetic field of $\HOOP=\Happl\cos{\theta}$.

Prior to acquiring images, the sample was tilted by \SI{-40}{\degree} with respect to the film normal in an applied magnetic field of about \SI{2}{\tesla} to align the in-plane magnetization of the sample toward the negative direction. Subsequently, \Happl{} was increased to zero in order to bring the in-plane magnetization to the negative remanence point. Images were then acquired at room temperature with the sample tilted at $\theta=+\SI{40}{\degree}$, beginning at the negative remanence point and increasing \Happl{} in steps up to \SI{2}{\tesla}. It is important to point out that, by tilting the sample by $+\SI{40}{\degree}$, the LTEM contrast is more enhanced than that of the zero tilt case (images not shown), which is consistent with the average magnetization angle measured by M\"ossbauer spectroscopy of about \SI{45}{\degree} from the film normal.

\Cref{fig:LTEM}(a) shows the magnetic domain structure at zero magnetic field, revealing that the size of the magnetic features is similar to the grain size (determined by the bright-field TEM images in \cref{fig:TEM}(a)). By increasing the applied magnetic field \Happl{} from zero to \SI{144}{\milli\tesla} (i.e., $\HIP=\SI{92.6}{mT}$ and $\HOOP=\SI{110.3}{mT}$), individual switching events can be seen (\cref{fig:LTEM}(a\textendash d)). The switching events are more clear in the corresponding difference images presented in \cref{fig:LTEM}(e,f). The small reversed areas are most likely noncollinear spin clusters arising from the competing ferromagnetic and antiferromagnetic interactions between the Fe atoms, as will be discussed in Sec.~\ref{sec:atomistic}.

At higher magnetic fields, up to \SI{2}{\tesla}, only a few individual switching events are observed (images not shown). Furthermore, the overall LTEM contrast is reduced, implying that the magnetization is predominantly oriented along the applied magnetic field direction. This is consistent with \cref{fig:MHcurves}(b), which shows that the magnetization of Ta$|$Ru$|$Fe$_{92}$Ru$_{8}$(20)$|$Ru is reversible at fields above about \SI{0.1}{\tesla}.
The individual reversal events can be observed in the video provided in the Supplemental Material \cite{supplemental}, which was recorded via LTEM of a sample area as an external magnetic field was increased from 96 to \SI{112}{\milli\tesla}.

\section{Discussion}
\subsection{Experimental Results}
\label{sec:exp_res}

The structural measurements, X-ray diffraction (XRD) and transmission electron microscopy (TEM), show that the studied films are polycrystalline with grains that are randomly oriented in the plane of the film. The grains with bcc crystal structure are textured along the growth direction, $\langle110\rangle$, and the grains with hcp crystal structure are textured along the $\langle001\rangle$ directions. For Ta$|$Ru$|$\xFe{}$|$Ru films with both bcc and hcp \xFe{}, rocking curve measurements show that the full-width-at-half-maximum is \SI{2.6}{\degree} or below.

The XRD and TEM measurements indicate that room-temperature \xFe{} transitions from bcc to hcp crystal structure as $x$ increases from \SI{\sim13}{\at} to \SI{\sim20}{\at}. 
From the TEM measurements, we also observed that the Fe$_{87}$Ru$_{13}$ and Fe$_{83}$Ru$_{17}$ films consist of grains having entirely hcp or entirely bcc lattice structure; furthermore, each film has indications of individual grains showing both phases.
STEM-EDXS analysis measurements of the Ta$|$Ru$|$\xFe{}$|$Ru structure reveal a uniform distribution of Ru throughout the FeRu films.

XRD measurements show that the lattice expansion with the addition of Ru to Fe is linear for up to \SI{8}{\at} of Ru in Fe, indicating that Ru atoms substitute Fe atoms in the bcc crystal structure. Out-of-plane and in-plane measurements of \xFe{} for $0\leq x\leq8$ indicate no difference in the lattice spacing for the in-plane and out-of-plane directions, 
suggesting that the cubic structure is not distorted by the addition of Ru.

Our M\"ossbauer spectroscopy measurements show no magnetic order in \xFe{} films with $x\geq19.5$, while films with $x\leq13$ contain more than $96\%$ of the magnetic phase. Combining these measurements with the structural measurements suggests that \hFeRu{} is paramagnetic and \bFeRu{} is magnetic.
In the absence of an external magnetic field, the angle of the average magnetization with respect to the film normal, \phiMS{}, of Ta$|$Ru$|$\xFe{}$|$Ru increases from
\SI{40 \pm2}{\degree} for $x=18.5$, to \SI{51 \pm1}{\degree} for $x=4$, and to \SI{68 \pm2}{\degree} for $x=0$. For a single layer of Fe, \phiMS{} increases to \SI{82\pm1}{\degree}. Since domain walls have out-of-plane magnetization components, the angle was also measured in the presence of a \SI{0.166}{T} external magnetic field to expel all magnetic domains in the film. With the field applied, \phiMS{} increased slightly to \SI{55 \pm1}{\degree} for $x=4$ and to \SI{85 \pm2}{\degree} for the single Fe layer. Therefore, magnetization in domain walls does not contribute significantly to the out-of-plane magnetization component observed.
Micromagnetic simulations presented in the Supplemental Material \cite{supplemental} support the presence of an out-of-plane magnetization component in our \SI{100}{nm}-thick single Fe layer with domains expelled. The simulations give an average angle between the magnetic moments in the film and the film normal of \SI{84.4\pm1.0}{\degree} in the presence of a \SI{0.05}{T} in-plane magnetic field.

The room-temperature $M(H)$ measurements of the \xFe{} films with the magnetic field applied parallel to the film plane show that $M_s$ initially decreases slowly with $x$ up to \SI{10}{\at}, which is predominantly due to Ru atoms substituting Fe atoms in the host bcc lattice. For $13\leq x\leq23$, $M_s$ sharply decreases to 0, which is mainly due to the magnetic phase transforming into the paramagnetic phase. On the other hand, the remanent magnetization $M_r$ sharply decreases with $x$ up to \SI{8}{\at} and decreases at a lower rate for larger $x$; the $M_r/M_s$ ratio follows the same trend as $M_r$. This trend correlates with the increase in the out-of-plane magnetization component observed by M\"ossbauer spectroscopy, i.e., the decrease in the angle \phiMS{} with respect to the film normal.
The $M_r/M_s$ measurements of the \xFe{} films at \SI{5}{K} are similar to those at room temperature except for $x\geq17$, for which $M_r/M_s$ increases due to the increase in the magnetic order of the structure.

The $M(H)$ measurements of the \xFe{} films with the field applied perpendicular to the plane measured a very small magnetization along the film normal at zero field for all measured compositions. This reveals that the majority of the out-of-plane magnetization components detected by M\"ossbauer spectroscopy for $x\leq18.5$ at zero field cancel each other out: the out-of-plane components must point in both directions along the film normal.
This is expected, as the demagnetizing field in the film is minimized when the net magnetization along the film normal is zero.

The MOKE microscopy measurements showed that magnetic domains are entirely expelled in the reversible part of the $M(H)$ loop. The field at which the $M(H)$ loop becomes reversible, $H_\text{rev}$, increases from $\SI{0.0037}{T}$ for the single Fe layer to \SI{0.13}{T} for Ta$|$Ru$|$Fe$_{92}$Ru$_{8}$$|$Ru. The normalized magnetization at these fields, $M(H_\text{rev})/M_s$, is about 0.96 for the single Fe layer and 0.59 for Ta$|$Ru$|$Fe$_{92}$Ru$_{8}$$|$Ru. The MOKE measurements also show that there is no stripe domain pattern present in our films to contribute to the large out-of-plane magnetization component observed within the film. 

For ferromagnetic films with non-interacting magnetic grains whose easy axes are randomly oriented in all directions, the 
$M_r/M_s$ ratio is 0.5 (see Supplemental Material \cite{supplemental}). If the easy axes of the magnetic grains are randomly oriented only in the plane of the film, with no out-of-plane components due to large in-plane magnetic anisotropy, $M_r/M_s$  increases to 0.64 (see Supplemental Material \cite{supplemental}). In cases where there is a strong direct exchange interaction between the grains, as in our films, the magnetic moments are pulled toward each other, significantly increasing $M_r/M_s$. This is corroborated by our pure Fe films, for which $M_r/M_s$ is 0.94. 
On the contrary, $M_r/M_s < 0.5$ for our \xFe{} films with $x \geq 8$, suggesting that a reduced interaction between magnetic grains cannot explain the low $M_r/M_s$ ratio in our RuFe films.

The field required to saturate the magnetic moments of ferromagnetic materials along the direction of the external field, the saturation field, can be calculated as $2K_1/M_s$, where $K_1$ is the first anisotropy constant and $M_s$ is the saturation magnetization of the ferromagnetic material \cite{coey2010magnetism}. Here, we assume $K_2 = K_3 = 0$ and neglect the demagnetizing fields. For pure Fe, 
$K_1 = \SI{48}{kJ/m^3}$ \cite{coey2010magnetism}, resulting in a saturation field of \SI{0.056}{T}. Since the magnetic phase of FeRu has a cubic crystal structure (bcc) and no structural distortion is observed for Fe$_{100-x}$Ru$_x$ films with $x \leq 8$, $K_1$ is small, resulting in a relatively small saturation field. The observed saturation field in our FeRu films is two orders of magnitude larger than that of Fe, suggesting that our films are not ferromagnetic (i.e., the magnetic interactions do not cause the magnetic moments of Fe atoms to align in the same direction).

Furthermore, our samples have a large $M_s$, exceeding $\SI{1.5}{MJ/m^3}$ for \xFe{} with $x \leq 8$, which induces a large in-plane shape anisotropy in our films. Due to the small magnetocrystalline anisotropy of the magnetic phase of FeRu, it is expected that the shape anisotropy will force the magnetization of the film to lie in the film plane. On the contrary, the average magnetization is aligned at an angle between \SI{39}{\degree} and \SI{50}{\degree} out of the film plane (i.e., \phiMS{} between \SI{51}{\degree} and \SI{40}{\degree}) for $x \geq 4$ at room temperature. Even in the presence of an external magnetic field large enough to expel all domains, the average magnetization was at an angle of at least \SI{35}{\degree} out of the film plane for $x \geq 4$. This clearly shows that magnetization in domain walls is not responsible for the observed out-of-plane magnetization component.
Thus, the out-of-plane magnetization component is another indication that our films are not ferromagnetic films.

The low $M_r/M_s$, high saturation field, and presence of an out-of-plane magnetization component observed in our FeRu films can be explained by noncollinear coupling between magnetic Fe atoms within the films, which gives rise to a nanogranular magnetic domain structure as observed by LTEM for a \xFe{} film with $x = 8$. The noncollinear coupling can arise from the competition between ferromagnetic coupling between neighbouring Fe atoms and antiferromagnetic coupling between Fe atoms separated by Ru atoms. This will be discussed in the following subsection.

\subsection{Atomistic model}
\label{sec:atomistic}

The recently-developed atomistic model \cite{abert2022origin} was successfully used to simulate the magnetic coupling of Co layers across FeRu layers. In the model, the crystal structure is simplified to simple cubic with a lattice constant of $a=\SI{0.25}{nm}$. Neighbouring Fe atoms couple ferromagnetically due to Heisenberg exchange interactions, while two Fe atoms separated by one or more Ru atoms couple antiferromagnetically. The model only includes ferromagnetic and antiferromagnetic couplings between pairs of Fe atoms along the three principal axes of the simple cubic unit cell. Further details about the atomistic model can be found in Ref.~\cite{abert2022origin} and Ref.~\cite{nunn2023controlling}.

\begin{figure}[tbph]
			\centering
		\includegraphics[width=8.6cm]{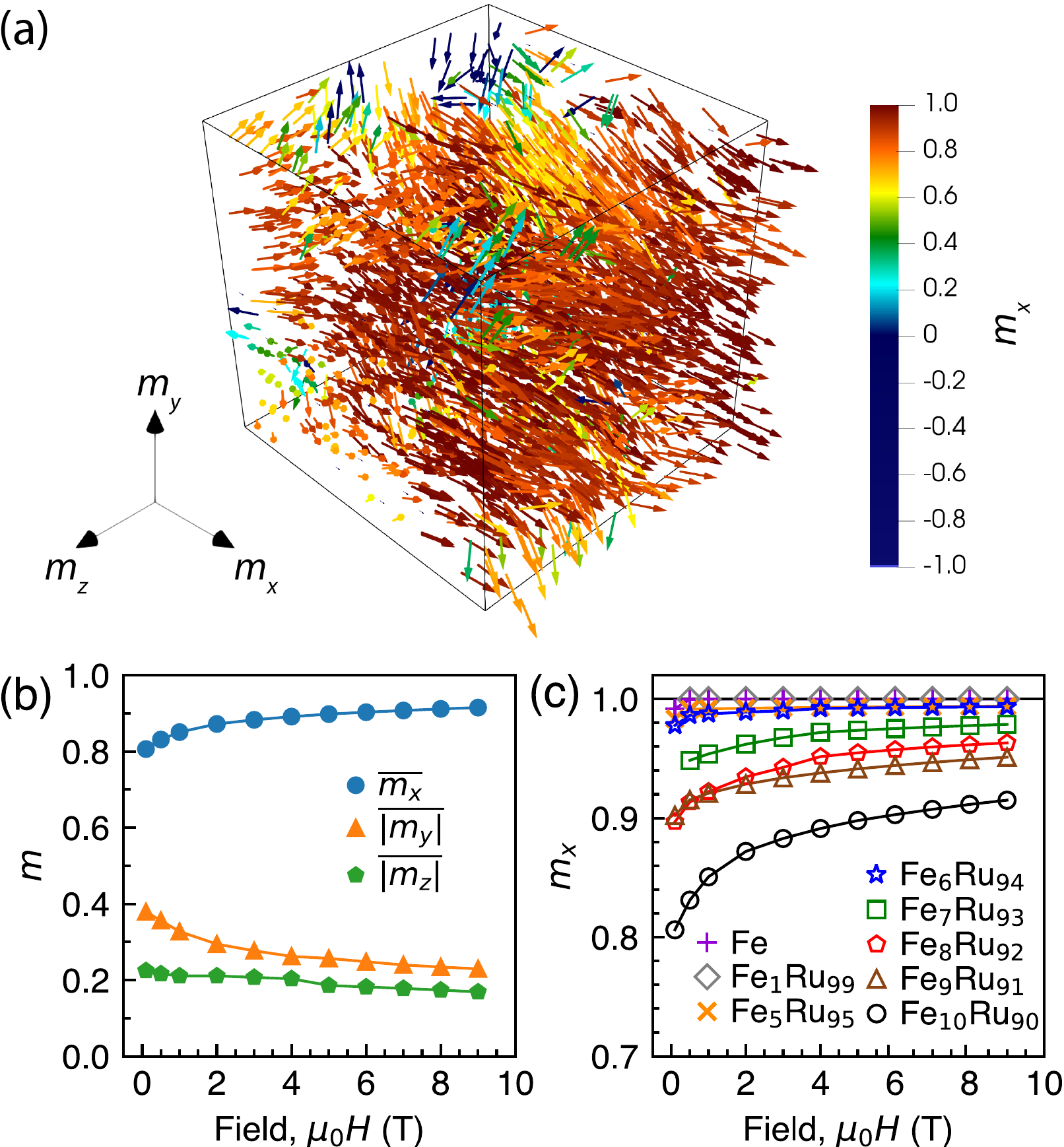}
            \caption{(a) Simulated magnetic moment distribution of an Fe$_{90}$Ru$_{10}$ film after magnetic field along $x$-axis is removed. (b) For the same film, average projection of normalized magnetic moments along $x$-axis (field direction) and average absolute value projections along $y$-axis and $z$-axis (film normal) as a function of applied magnetic field. (c) Simulated normalized $M(H)$ dependence of \xFe{} films ($0\leq x\leq 10$).}
 		\label{fig:sim}
 	\end{figure}

We used the atomistic model to simulate \xFe{} films for $0\leq x\leq10$, using a saturation magnetization of $M_s=\SI{1700}{kA/m}$ to obtain the dipole moment of the Fe atoms and an exchange stiffness of $A_\text{ex}=\SI{21}{pJ/m}$ to obtain the Heisenberg exchange constant between the Fe atoms. The antiferromagnetic coupling strengths of Fe across one, two, and three Ru atoms are \SI{-3.13e-21}, \SI{-1.14e-21}, and \SI{-4.43e-22}{J}, respectively.
To account for the large shape anisotropy in our \xFe{} films, we introduced a demagnetizing field, with the easy-plane shape anisotropy being $K_\text{shape}=-\mu_0 M_s^2/2$.
\Cref{fig:sim}(a) shows the magnetic moment distribution in the Fe$_{90}$Ru$_{10}$ film after a large magnetic field is applied along the $x$-axis, parallel to the film plane, and then reduced to zero.
The complex configuration is caused by the competition between the ferromagnetic and antiferromagnetic couplings between the Fe atoms.
The vast majority of magnetic moments have a positive projection along the magnetic field direction after the field is removed, and for this reason, the colour of the magnetic moments with a negative projection is kept the same in \cref{fig:sim}(a).

From the magnetic moment distribution of the Fe$_{90}$Ru$_{10}$ film, we calculated the average projection of the normalized magnetic moments on the $x$-axis (the field direction), the $y$-axis, and the $z$-axis (the film normal) as a function of the external magnetic field, as shown in \cref{fig:sim}(b). Since the magnetic field is applied along the $x$-axis, the magnetic moments preferentially point in the positive $x$ direction, as expected; in fact, $\overline{m_x}\geq0.8$ for all fields greater than or equal to \SI{0.1}{\tesla}.
However, the magnetic moments have no analogous preferential direction along the $y$-axis or $z$-axis, causing the average projection on these two axes to be zero. For this reason, we have plotted the average of the absolute value of the normalized magnetic moment projections along the $y$ and $z$ directions, $\overline{|m_y|}$ and $\overline{|m_z|}$, respectively.
The absolute value projection along the $z$ direction is similar to the magnetization detected by M\"ossbauer spectroscopy.
\Cref{fig:sim}(a,b) show that there is a substantial magnetization component in the $z$ direction, perpendicular to the film plane. This can explain the out-of-plane magnetization component observed by M\"ossbauer spectroscopy.

In \cref{fig:sim}(c), the average projection of the normalized magnetic moments along the field direction, $\overline{m_x}$, is plotted as a function of external magnetic field for \SI{100}{nm}-thick \xFe{} films in the concentration range $0\leq x\leq10$. These curves correspond with our normalized $M(H)$ curves measured by VSM.
\Cref{fig:sim}(c) shows a decreasing trend in $M_r/M_s$ with increasing Ru concentration in Fe that agrees with our measurements in \cref{fig:MHcurves,fig:saturationloop}. Additionally, \cref{fig:sim}(c) shows that a large field is required to fully saturate the films, as shown in \cref{fig:saturationloop}(a). However, the simulated decrease in $M_r/M_s$ is much slower than the experimental decrease in $M_r/M_s$, and the saturation fields are much larger.

\section{Conclusion}

In summary, we have studied the structural and magnetic properties of sputter-deposited Fe and \xFe{} films by means of X-ray diffraction (XRD), transmission electron microscopy (TEM), M\"ossbauer spectroscopy, vibrating sample magnetometry (VSM), magneto-optical Kerr effect (MOKE) microscopy, and and Lorentz transmission electron microscopy (LTEM). All studied films are polycrystalline, and the \xFe{} films, which were grown on a Ta$|$Ru seed layer, exhibit strong texture along the growth direction. The crystal structure of the \xFe{} films is predominantly bcc for $x<13$ and undergoes a gradual transition to hcp in the concentration range $13 \lesssim x \lesssim 20$.
Within this transitional range, the films consist of grains that have fully hcp, fully bcc, and a mixture of hcp and bcc lattice structures.

While Fe has ferromagnetic order, the addition of Ru introduces a noncollinear magnetic coupling within \bFeRu{}. The noncollinear coupling is attributed to the competition between ferromagnetic and antiferromagnetic coupling in the film: ferromagnetic coupling between neighbouring Fe atoms and antiferromagnetic coupling between Fe atoms separated by Ru atoms. On the other hand, \hFeRu{} was found to be paramagnetic at room temperature.

Thin-film FeRu was recently shown to mediate a large noncollinear interlayer coupling between two hcp ferromagnetic layers, for Ru concentrations from 20 to \SI{38}{\at} \cite{nunn2020control}. Here we show that, for the FeRu composition range of interest for noncollinear coupling, FeRu maintains hcp crystal structure. This is important for the epitaxial growth of these noncollinearly-coupled trilayer structures.

Magnetic multilayer structures used for applications, including magnetic sensors and magnetic solid state memory, make extensive use of both Fe and Ru layers. Here we showed that the presence of a small amount of Ru in Fe can induce noncollinear magnetization alignment between Fe atoms. This can strongly affect the performance of these devices.

\section{Acknowledgement}

The authors thank Sabri Koraltan for useful discussions, Romy Aniol for TEM specimen preparation, Afan Terko for plotting \cref{fig:sim}, and Dr. Jan Rhensius and Simon Josephy from QZabre for useful discussions. We acknowledge financial support from the Natural Sciences and Engineering Research Council of Canada (NSERC), Fonds Qu\'{e}b\'{e}cois de la Recherche sur la Nature et les Technologies (FRQNT), and the Austrian Science Fund (FWF) P 34671.
The use of the HZDR Ion Beam Center TEM facilities and the funding of TEM Talos by the German Federal Ministry of Education and Research (BMBF; Grant No. 03SF0451) in the framework of HEMCP are also acknowledged.

%


\begin{thebibliography}{32}
\bibitem [{sup()}]{supplemental}%
  \BibitemOpen
  \bibinfo {title} {See {S}upplemental {M}aterial at [url will be inserted by
  publisher] for additional {TEM} images, micromagnetic simulations of an {F}e
  film, calculations of the {$M_r/M_s$} ratio for a ferromagnetic film with
  non-interacting magnetic grains, and a {L}orentz transmission electron
  microscopy video of switching events in an {F}e$_{92}${R}u$_8$
  film.}\BibitemShut {Stop}%
\end{thebibliography}

\end{document}